\documentclass{aa}  

\usepackage{graphicx}
\usepackage[export]{adjustbox}
\usepackage{comment}
\usepackage[varg]{txfonts}
\usepackage{xcolor}
\usepackage{upgreek}
\usepackage[normalem]{ulem}

\renewcommand{\vector}[1]{\ensuremath{\pmb{#1}}}
\renewcommand{\div}{\ensuremath{-}}

\newcommand{\Msun}{\ensuremath{\, M_\odot}}

\newcommand{\gcm}{\ensuremath{\,\rm g\, cm^{-2}}}

\DeclareRobustCommand{\uppartial}{\text{\rotatebox[origin=t]{20}{\scalebox{0.95}[1]{$\partial$}}}\hspace{-1pt}}
\newcommand{\pardir}[2]{\ensuremath{\frac{\uppartial #2}{\uppartial #1} }}

\newcommand{\ppardir}[2]{\ensuremath{\frac{\uppartial }{\uppartial #1} \left( #2\right)}}
 
\renewcommand{\i}{\ensuremath{{\rm i}}}
\newcommand{\diff}{\ensuremath{{\rm d}}}

\begin{document} 
 
\title{Kilohertz quasi-periodic oscillations  from neutron star spreading layers}
   
\titlerunning{Kilohertz QPOs from neutron star spreading layers}
 
\authorrunning{P. Abolmasov et al.}
 
\author{Pavel Abolmasov
          \inst{1,2}
          \and
          Joonas N\"attil\"a\inst{3,4,5}
          \and
          Juri Poutanen\inst{1,5,6}
          }

\institute{Department of Physics and Astronomy, FI-20014 University of Turku, Finland\\
     \email{pavel.abolmasov@gmail.com}
     \and
     Sternberg Astronomical Institute, Moscow State University,
     Universitetsky pr. 13,  119234 Moscow, Russia
    \and 
    Physics Department and Columbia Astrophysics Laboratory, Columbia University, 538 West 120th Street, New York, NY 10027, USA
    \and 
    Center for Computational Astrophysics, Flatiron Institute, 162 Fifth Avenue, New York, NY 10010, USA  
    \and
    Nordita, KTH Royal Institute of Technology and Stockholm University, Roslagstullsbacken 23, SE-10691 Stockholm, Sweden 
    \and
    Space Research Institute of the Russian Academy of Sciences, Profsoyuznaya str. 84/32, 117997 Moscow, Russia
   }

   \date{Received ; accepted }

  \abstract
  {When the accretion disc around a weakly magnetised neutron star (NS) meets the stellar surface, it should brake down to match the rotation of the NS, forming a boundary layer. As the mechanisms potentially responsible for this braking are apparently inefficient, it is reasonable to consider this layer as a spreading layer (SL) with negligible radial extent and structure. 
  We perform hydrodynamical 2D spectral simulations of an SL, considering the disc as a source of matter and angular momentum. Interaction of new, rapidly rotating matter with the pre-existing, relatively slow material co-rotating with the star leads to instabilities capable of transferring angular momentum and creating variability on dynamical timescales. 
  For small accretion rates, we find that the SL is unstable for heating instability that disrupts the initial latitudinal symmetry and produces large deviations between the two hemispheres.
  This instability also results in breaking of the axial symmetry as coherent flow structures are formed and escape from the SL intermittently.
  At enhanced accretion rates, the SL is prone to shearing instability and acts as a source of oblique waves that propagate towards the poles, leading to patterns that again break the axial symmetry.
  We compute artificial light curves of an SL viewed at different inclination angles. Most of the simulated light curves show oscillations at frequencies close to 1~kHz. 
  We interpret these oscillations as inertial modes excited by shear instabilities near the boundary of the SL. 
  Their frequencies, dependence on flux, and amplitude variations 
  can explain the high-frequency pair quasi-periodic oscillations observed in many low-mass X-ray binaries. }

   \maketitle
%

\section{Introduction}

For a non-magnetised accretor with a fluid or solid surface, non-relativistic
disc accretion
releases only about half of the gravitational binding energy. The other half is stored as
kinetic energy of the flow (see e.g. \citealt{SS00}). As the accretor is unlikely to
rotate close to its breakup limit, a greater  proportion of this energy will still
dissipate close to the surface of the accretor in what is called a boundary
layer (BL). There is no commonly accepted view of a BL; even its basic
geometry is uncertain.

Boundary layers are expected to appear during disc accretion onto stars and
compact objects with relatively weak magnetic fields that are incapable of creating a
magnetosphere. For neutron stars (NS), this happens for a surface magnetic field
$B\lesssim 10^8$~G if the accretion rate is approximately  Eddington. 
Lower mass accretion rates set a stronger limit for the magnetic field that is proportional to the square root of the accretion rate (see e.g. \citealt{LPP}).
This case is relevant for old NSs in low-mass X-ray binaries (LMXBs). 
In particular, the BL apparently plays an important role in the so-called Z and atoll sources (as classified by \citealt{HK89}). 
Combined spectral and timing observations of LMXBs allow the contribution of the BL to be separated from that of the accretion disc \citep{revnivtsev03, revnivtsev06}. Emission of the boundary layer is hotter, with a colour temperature of about 2.5~keV. The BL spectral component is more variable
than the disc on short, dynamical timescales. In particular, the highest-frequency, kilohertz quasi-periodic oscillations (kHz QPOs; see e.g. \citealt{vdKlis_review}) can be interpreted as some type of BL activity. This is supported by the fact that, while the properties of all the low-frequency QPO types are quite similar in NS and black-hole LMXBs, kHz QPOs behave in a profoundly different way for NS sources \citep{motta}.
 It appears natural to attribute this  difference to the existence of a solid surface and a BL.
 
These QPOs are often observed in pairs, and the distance between peaks is either close to the frequency of burst oscillations (itself well consistent
with the spin frequency of the NS; see \citealt{vdKlis_review}) or to one
half of this value \citep{mendez98, wijnands03}. As the observational data accumulate, the picture becomes more complicated, favouring rather a universal, sometimes variable frequency difference of $\Delta f \sim 300$~Hz \citep{MB07}, close to but not equal or proportional to the spin frequency. 
Normally, there are only two kHz peaks present. 
One explanation is a bright spot rotating at a frequency of the order of Keplerian frequency (for a conventional NS with a mass of $1.5\Msun$, and radius 12~km, linear Keplerian frequency is $f_{\rm K} \simeq 1.5$~kHz) and producing one peak due to visibility effects and the other due to interaction with non-axisymmetric structures at the surface of the NS. This interpretation is known as a beat-frequency model (first apparently proposed by \citealt{lamb85} in the context of a different QPO type) and implies strict equality between the difference in the frequencies of the two QPOs and the spin frequency of the NS. 
Also, different types of beat-frequency models that do not involve a BL were proposed to explain the properties of kHz QPOs. 
The best known are the magnetospheric beat-frequency model \citep{psaltis98}, considering LMXBs as magnetised accretors with very compact magnetospheres, and the sonic-point beat frequency model \citep{miller98}. The latter relies on the strong gravity effects that for a conventional NS place the last stable Keplerian orbit at a radius somewhat larger than the radius of the star. Both models require some mechanism generating narrow-band variability at the local Keplerian frequency. Resonances between Keplerian, epicyclic, and vertical epicyclic frequencies emerging in general relativistic solutions are apparently a good explanation for the kHz QPOs in black hole systems \citep{KA01b,kluzniak04}, but predict a fixed ratio of 2:3 for the peak frequencies. 
In NS systems, the frequency ratio varies in rather broad limits, and 2:3 is only a crude approximation. 
The caveats of the conventional resonance model were considered by \citet{rebusco08}. 

Quasi-periodic oscillation frequencies shift with time, remaining correlated with the observed flux on short timescales (hours and less) and uncorrelated on longer timescales. 
This creates parallel tracks in the flux versus QPO-frequency plot \citep{mendez99, klis01},  suggesting that a BL possesses a characteristic correlation timescales much longer than even the viscous timescales in the disc (on which the mass accretion rate varies). 
It is difficult to suggest a way in which the accretion disc could produce the parallel tracks, as its variability is governed essentially by a single parameter: the mass accretion rate. 
If the BL is weakly coupled with the surface of the star, it is a good candidate for such an `integrator'. 
The rich phenomenology of kHz QPOs is a potentially important source of information about the NS itself and the physics of the flows close to its surface.
However, besides the numerous observational clues and the variety of existing models, little is known about the mechanisms and exact relations between the quantities. 

The classical approach to the BL considers the flow as some part of the disc \citep{pringle77, PN95} where the rotation velocity deviates strongly from Keplerian rotation and matches the rotation rate of the star at the inner edge. 
Strongly non-Keplerian rotation means that the approximations used as the basis for the thin disc approach are no longer valid, and the radial structure of the flow is strongly affected by the radial pressure gradient. 
Another problem is the efficiency of angular momentum exchange in the BL. 
In a hot (ionized) accretion disc with a rotation law close to Keplerian, there is an outward angular momentum transfer
provided by the magnetohydrodynamic (MHD) turbulence generated by magneto-rotational instability (MRI; introduced in the astrophysical context by \citealt{MRI}), which is only operational when  the angular velocity decreases with the cylindrical radial coordinate. For a BL, the rotation profile does not in general fulfil the necessary condition for MRI.
It is unclear which physical mechanisms are responsible for the angular
momentum exchange between the accreted matter and the surface of the star. 
In practically any possible BL model, the Rayleigh stability criterion \citep{rincon07} is satisfied. 
Hydrodynamic turbulence is still produced for large enough Reynolds numbers (see \citealt{ZR18} for a detailed analysis), but the amplitudes of turbulent motions are apparently insufficient for efficient angular momentum transfer. 
The very existence of the extremely long correlation timescale mentioned above suggests that irrespective of the mechanism of angular momentum transfer, it is extremely slow and inefficient.

A good candidate for such a mechanism is supersonic shear instability at
the interface between the star and the BL \citep{BRS13, HK15, philippov16, BQ18}. 
Unlike the classical subsonic Kelvin-Helmholtz instability, oblique waves rather
than vortices are generated. Moving in a shear velocity flow, the waves create
Reynolds stress and thus provide effective non-local
viscosity not only in the BL but also in the accretion disc. 
Most of the numerical studies mentioned above considered a two-dimensional problem either in the
equatorial plane or at a fixed latitude on a conical surface, as did \citet{philippov16}. 
\citet{BQ18} considered a three-dimensional local MHD problem with a fixed equation of state, mainly addressing the structure of the flow in the equatorial plane.
Depending on the particular simulation setup, the contribution of the wave-mediated angular momentum transfer varies from negligibly small to somewhat comparable to that of the MHD turbulence generated in the disc. 

If the angular momentum exchange rate between the accreting
matter and the material of the star is smaller than the angular momentum supply from the disc,
rapidly rotating matter would accumulate
on the surface, pushed to higher latitudes by pressure gradient. The radial
dimension of such a flow, as well as of a conventional BL, is second order in relative disc thickness, much smaller
than its vertical (along the polar angle) extent \citep{PS86}. Consequently, one can treat the flow as two-dimensional (2D) on
the surface of the NS fed by matter and angular momentum injection from the
disc. This approach, known as the \emph{spreading layer} (SL) approach, was introduced by
\citet{IS99} and further developed in \citet{SP06} and \citet{IS10}. For the case of accreting
white dwarfs, this model was considered by \citet{PB04a} and used to explain a
certain type of QPO observed in cataclysmic variables, the so-called
dwarf-nova oscillations (DNOs; \citealt{PB04b}).
The angular momentum transfer within the SL depends on the dynamics of the flow itself as well as existing oscillation and instability modes. 
It is quite probable that certain hydrodynamical phenomena will provide an efficient way to transfer momentum within the SL and thus define its internal dynamics. 
In this paper, we try to address the issue of angular momentum transfer within the layer using numerical hydrodynamical simulations.

Two-dimensional hydrodynamics on a rotating sphere is an important subject in geophysics and astrophysics, as many spherical bodies, including planets,
stars, and NSs, have fluid atmospheres.
Vertically integrated equations of hydrodynamics lead to the system of \emph{shallow water} equations (see e.g.~\citealt{vreug}), normally used in geophysics for weather
forecasts in combination with spectral methods \citep{jchien}.
Spectral methods provide much higher accuracy than finite-difference methods
on an equally fine grid, and are quite robust and stable for subsonic
flows. 
However, rotation in a SL is almost always supersonic, which makes the
flow compressible and its simulations potentially prone to numerical
instabilities. 
This makes numerical simulations of spreading layers technically challenging.
On the other hand, using spherical harmonics is natural in spherical coordinates and allows  the singularity of the spherical grid near its axis to be avoided.
We provide our full simulation code \textsc{SLayer} as open-source software
\footnote{\url{https://github.com/pabolmasov/SLayer}}.

The paper is organised as follows. In Sect.~\ref{sec:model}, we formulate
the physical problem and derive all the basic equations. Results of the simulations are given in  Sect.~\ref{sec:res}. Applications and
limitations of the model are discussed in Sect.~\ref{sec:disc}. We conclude in Sect.~\ref{sec:conc}. A detailed description of the numerical techniques is given in Appendix~\ref{sec:numerics}. 

\section{Physical model}
\label{sec:model}

\subsection{Scales and dimensionless quantities}

Natural timescale in the vicinity of a relativistic object of mass $M$ is
\begin{equation}\label{E:set:timescale}
  t_{\rm g} = \frac{GM}{c^3} \simeq 7\times 10^{-6}  \frac{M}{1.4\Msun}\ \mbox{s},
\end{equation}
which is the approximate light-crossing or dynamical timescale at the event horizon. 
The corresponding radius is
\begin{equation}\label{E:set:radius}
  R_{\rm g} = \frac{GM}{c^2} \simeq 2 \frac{M}{1.4\Msun}\  \mbox{km}.
\end{equation}
The radius of the NS is taken as $R_* \simeq 12\,\mbox{km} \simeq 6 R_{\rm g}$ assuming a mass of $1.4\Msun$ (see e.g. the estimates of masses and radii in \citealt{ML16,nattila17}).

All the geometrical and kinematic quantities are naturally normalised by combinations of these spatial and temporal scales.
In particular, velocities in units of the speed of light $c$ are used.
We use physical quantities (g\,cm$^{-2}$) for $\Sigma$ and 
internally normalise the surface density by an arbitrary scale set either to $10^4$
or to $10^8\gcm$ for the simulations presented in this paper. 
Evidence for a very long correlation timescale $t_{\rm corr}$ suggests a characteristic value of
\begin{equation}\label{E:set:sigma}
  \Sigma_{\rm ch} \sim \frac{\dot{M} t_{\rm corr}}{4\uppi R^2 }\simeq 2\times 10^8
  \frac{\dot{M}}{10^{18}\ \mbox{g\,s}^{-1}} \frac{t_{\rm corr}}{1\,{\rm hour}} \left(
  \frac{12\ \mbox{km}}{R_*}\right)^2 \gcm.
\end{equation}
The physical meaning of this value is the mean surface density if $t_{\rm corr}$ is a characteristic time of mass depletion or replenishment in the SL. The vertical optical depth of the layer is simply $\varkappa \Sigma$, where $\varkappa $ is opacity. Here, we set $\varkappa$ to Thomson electron scattering opacity for Solar metallicity, $\varkappa_{\rm T} \simeq 0.34{\rm \, cm^2\, g^{-1}}$.
  
The problem has a complex hierarchy of timescales, starting with the dynamical scale, which is normally the smallest. 
The characteristic dynamical timescale is the Keplerian period near the surface of the NS: 
\begin{equation}
    t_{\rm dyn} = 2\uppi \sqrt{\frac{R_*^3}{GM_*}} \simeq 6\times 10^{-4} \left( \frac{R_*}{12\ \mbox{km}}\right)^{3/2} \left( \frac{M_*}{1.4\Msun}\right)^{-1/2}  \ \mbox{s}.
\end{equation}
The local thermal timescale depends on the effective and internal temperatures. 
For LMXBs, in Z and brighter atoll states, mass accretion rates
span the range $10^{15}\div 10^{18}\ \mbox{g\,s}^{-1}$ ($10^{-11}\div 10^{-8}\Msun\ \mbox{yr}^{-1}$) which implies effective temperatures of the order 
\begin{eqnarray}\label{E:set:teffscale}
 \displaystyle T_{\rm eff} &\simeq& \left( \frac{GM\dot{M}}{8\uppi \sigma_{\rm SB} R_*^3}
  \right)^{1/4} \nonumber \\
  \displaystyle &\simeq & 1.4\left(\frac{M}{1.4\Msun}\right)^{1/4} \left(\frac{R_*}{12{\rm km}}\right)^{-3/4} \left( \frac{\dot{M}}{10^{18}\ \mbox{g\,s}^{-1}}\right)^{1/4} \ \mbox{keV}. 
\end{eqnarray}
As half of the accretion power is emitted by the disc, we use $8\pi$ in the denominator for this estimate.
Internal temperature, $T_{\rm in}$, is about a factor $(\varkappa \Sigma)^{1/4}\sim 100$ times larger if $\Sigma \sim 10^8\gcm$ is taken as a
representative value. The thermal timescale is then easy to estimate, separating contributions of the radiation and gas pressure. If $\beta$ is the gas pressure fraction, 
\begin{equation}\label{E:set:tthscale}
  \begin{array}{l}
    \displaystyle  t_{\rm thermal, \ gas} \sim \frac{E}{Q^-} \sim 
    \frac{12}{5} \frac{1-\beta/2}{\beta} \frac{2kT_{\rm in}\Sigma }{m_{\rm p} \sigma T_{\rm eff}^4} \\
\qquad{}  \displaystyle   \sim 40  \frac{1-\beta/2}{\beta} \frac{\Sigma}{10^8\gcm} \left( \frac{T_{\rm
      in}}{100\,\mbox{keV}}\right) \left( \frac{T_{\rm eff}}{1\,\mbox{keV}}\right)^{-4}
       \ \mbox{s},\\
  \end{array}
\end{equation}
where $Q^-$ represents the energy lost via radiation (introduced more rigorously in Sect.~\ref{sec:energy}). 
Dependence on the gas pressure fraction follows from the relations between the vertically integrated and local quantities derived below in Sect.~\ref{sec:vert}. 
As the gas becomes radiation-pressure dominated, $\beta$ approaches zero, and the thermal timescale becomes longer. As we show below (Eq.~\ref{E:vert:beta}), the gas pressure ratio is related to the Eddington factor 
$\varkappa Q^- / c g_{\rm eff} = 1-\beta$. 
Violation of the local Eddington limit in our model is impossible as it makes the effective gravity negative, and the material of the layer unbound. On the other hand, the amount of internal energy stored in the layer is essentially unlimited ($E\propto \beta^{-1}$).
However, for very small $\beta $, the vertical thickness of the layer (see Eq.~\ref{E:vert:H}) becomes comparable to $R_*$ and thus limits the thermal energy stored in the SL.
For $T_{\rm in} = 100\,\mbox{keV}$, the minimal possible gas pressure fraction is about $\beta_{\rm min} \simeq 10^{-4}$, which corresponds to a thermal timescale longer than a day, meaning that very rapid accretion effectively runs in a radiatively inefficient regime. 
If energy release is close to equilibrium with radiation losses, the effective temperature does not change significantly, and most of the variations of the thermal timescale are related to the energy stored by gas and trapped radiation. 
Thermal and dynamical timescales become comparable if the surface density is small, $\Sigma \lesssim 10^3\gcm$. 

The challenge of SL simulations for the case of LMXBs is in the relatively long accretion timescale. 
While phenomena like kHz QPOs manifest themselves on
dynamical timescales of milliseconds, the accretion rate is relatively stable at the scales of
minutes to hours (viscous times of the inner disc), and the putative time of
mass growth and depletion in a SL is apparently of the same order. However, these timescales are much longer than both the thermal and dynamical timescales.

Hence, all the dynamical  and thermal-timescale phenomena we intend to
consider appear in fact in a quasi-stationary SL where mass accreted during
the considered timescales is negligibly small.
However, to reach such an equilibrium state, one needs to simulate either an episode of much more rapid
accretion or, alternatively, accretion atop a much thinner atmosphere, where the surface
densities of pre-existing and newly accreted material would be comparable on
a reasonable timescale of the simulation run. We try both approaches. 

\subsection{Geometry and mass conservation}\label{sec:mass}

\begin{figure}
\adjincludegraphics[width=\columnwidth,trim={0 {0.\width} 0 0},clip]{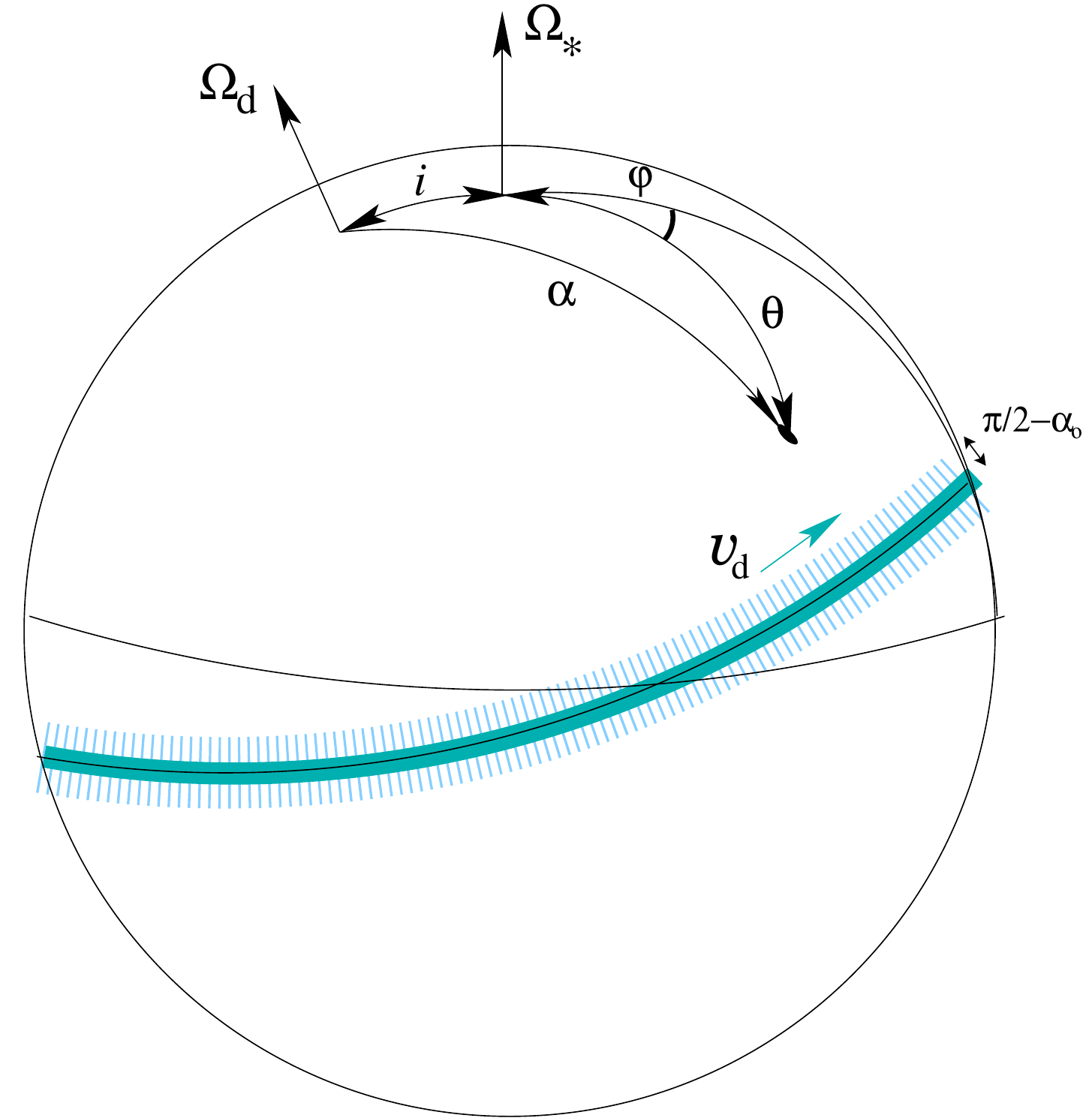}
\caption{Illustration of the model geometry. 
The tilted blue arc near the equator shows the source of mass and momentum. 
The spin axis of the star, marked with $\Omega_*$, is inclined with respect to the disc axis ($\Omega_{\rm d}$), by an angle $i$. }
\label{fig:spheresketch}
\end{figure}

Continuity equation for density $\rho$ and velocity $\vector{v}$ in the most
general Newtonian form is
\begin{equation}\label{E:mass:rho}
\pardir{t}{\rho} = - \nabla \cdot (\rho \vector{v}) + \mbox{source/sink terms}.
\end{equation}
Integration over $r$ yields
\begin{equation}\label{E:mass:sigma}
\pardir{t}{\Sigma} = - \nabla \cdot (\Sigma \vector{v}) + S^+ - S^- ,
\end{equation}
where $S^\pm$ are the source and sink terms for surface density. The source term is set explicitly
as an inclined belt (to account for the case where the disc is inclined with respect to the rotation axis of the NS),
\begin{equation}\label{E:mass:source}
S^+ = S^+_{\rm norm}e^{-(\cos\alpha/\cos\alpha_0)^2/2},
\end{equation}
where $\alpha$ is the angular distance from the direction of the adopted disc
rotation axis,
\begin{equation}\label{E:mass:sina}
\cos \alpha = \cos \theta \cos i+\sin \theta \sin i \cos \varphi,
\end{equation}
where disc inclination $i$ with respect to the spin axis of the star sets the direction of the symmetry and rotation axis of the source (see Fig.~\ref{fig:spheresketch}). 
Here, we use a spherical coordinate system consisting of radial coordinate $r$,
co-latitude $\theta$, and longitude $\varphi$. 
Instead of $\theta$, latitude $\lambda = \uppi/2-\theta$ is  used below for visualisation. The system is aligned with the rotation of the star, but is itself non-rotating (inertial). 
The SL is considered thin in the radial direction, making $r,$ in a sense, a vertical coordinate, along which hydrostatic equilibrium is assumed to hold. See Sect.~\ref{sec:vert} for more details. 

Adding a sink allows to limit the growth of the mass of the SL and approach a quasi-stationary state. 
Matter existing long enough on the surface of the NS should adopt its velocity and physical properties. 
As our model
adopts a simplified vertical structure, adding a sink allows to draw an
effective boundary between the SL and the material of the NS.
In this paper, we ignore the sink term, but include it in the equations for future use in the form
\begin{equation}\label{E:mass:sink}
  S^- = \frac{\Sigma}{t_{\rm depl}},
\end{equation}
where $t_{\rm depl}$ is a depletion timescale set explicitly. 
This form is valuable for its simplicity and allows us to study the
role of the mass of the SL by considering a gradual depletion regime (mass
accretion is switched off but the sink is not) and the steady-state spreading
regime by turning simultaneously on both the source and the sink.

We also use a tracer quantity $a$ to separate the contributions of the
pre-existing and newly accreted material. It is assumed to be a passive scalar
quantity transferred with the flow and initially equal to zero. The source of
this quantity is designed in such a way as to reproduce the evolving fraction of newly accreted matter,
\begin{equation}\label{E:mass:accflag}
\displaystyle  \pardir{t}{a} + (\vector{v} \cdot \nabla) a = (1-a)
\frac{S^+}{\Sigma}.
\end{equation}

\subsection{Vertical structure}
\label{sec:vert}

In the vertical (radial) direction, SL is supported by thermal (gas and radiation) pressure together with the relevant centrifugal force component. 
As we consider the vertical extent of the SL to be infinitely small, the timescale of vertical dynamical relaxation should also be small, and therefore we neglect the effect of radial velocity in momentum and energy equations.
This allows us to write down a hydrostatic balance equation
\begin{equation}\label{E:vert:p}
\frac{1}{\rho}\pardir{r}{p} = -g_{\rm eff} = -\frac{GM}{r^2} +
\frac{v_\theta^2+v_\varphi^2}{r},
\end{equation}
which needs to be supplemented by another equation to calculate
the vertical profiles of pressure and density  simultaneously. 
Let us assume, following \citet{IS99} and \citet{SP06}, that the heat is released at the bottom of the SL, and the optical depth is high enough to use the radiation diffusion approximation. 
Hence, radiation flux $F$ is constant with height
\begin{equation}\label{E:vert:pr}
\displaystyle F = - \frac{c}{\varkappa \rho} \pardir{r}{p_{\rm rad}},
\end{equation}
where $p_{\rm rad}$ is radiation pressure. Total pressure is assumed to be contributed by $p_{\rm rad}$ and gas pressure $p_{\rm gas}$.
Together, equations (\ref{E:vert:p}) and (\ref{E:vert:pr}) imply a constant pressure ratio $\beta = p_{\rm gas} / p$ as long as opacity is constant with height. 
Hence, gas, radiation, and total pressure scale with each other, and the gas-to-total pressure ratio equals
\begin{equation}\label{E:vert:beta}
\displaystyle \beta = 1- \frac{\varkappa F }{cg_{\rm eff}}.
\end{equation}
Proportionality of pressures also implies $p \propto \rho T \propto T^4$, which
leads to $p\propto \rho^{4/3}$, an effectively polytropic law. This implies a relation between the pressure $p_0$ and density $\rho_0$ at the
bottom of the SL and the corresponding vertically integrated quantities, pressure $\Pi = \int p dr$ and surface density $\Sigma = \int \rho dr$, 
\begin{equation}\label{E:vert:pisigma}
\displaystyle \Pi = \frac{4}{5} \frac{p_0}{\rho_0} \Sigma.
\end{equation}
Constancy of $\beta$ allows also to link surface energy density $E = \int \varepsilon dr$ (where $\varepsilon$ is the volumetric
energy density) and integrated pressure $\Pi$ with each other. 
Local energy density may be expressed as
\begin{equation}\label{E:vert:eps}
\displaystyle \varepsilon = 3\left(1-\frac{\beta}{2}\right) p,
\end{equation}
which implies an identical relation for the vertically integrated quantities
\begin{equation}\label{E:vert:Eint}
\displaystyle E = 3\left(1-\frac{\beta}{2}\right) \Pi.
\end{equation}
The pressure ratio itself  may be found as a function of $E$ (or $\Pi$) and
$\Sigma$. At the bottom of the layer,
\begin{equation}\label{E:vert:bottombeta}
\beta = \frac{\rho_0 kT_0}{mp_0} = \frac{k}{m}
\frac{\rho_0}{p_0}\left(\frac{3}{4}\frac{c}{\sigma_{\rm SB}} \left(1-\beta
  \right) p_0\right)^{1/4},
\end{equation}
where $m \simeq 0.6m_{\rm p}$ is the mean mass of a massive particle,
which allows us to solve implicitly for $\beta$, taking into account the
expression for $p_0 = g_{\rm eff} \Sigma$ arising as a solution
to~(\ref{E:vert:p}) and equation (\ref{E:vert:Eint}) and substituting them into (\ref{E:vert:bottombeta}):
\begin{equation}\label{E:vert:ebeta}
\displaystyle \frac{\beta}{\left(1-\beta\right)^{1/4}\left(1-\beta/2\right)} = \frac{12}{5} \frac{k}{m}
\left(\frac{3}{4}\frac{c}{\sigma_{\rm SB}} g_{\rm eff} \Sigma \right)^{1/4}
\frac{\Sigma}{E} .
\end{equation}
The thickness of the layer may be found as the solution of hydrostatic equation, taking into account the constancy of $\beta$ (see also equation 32 in \citealt{SP06}) as
\begin{equation}\label{E:vert:H}
    H = \frac{5\Pi}{g_{\rm eff}\Sigma} = \frac{5}{\beta} \frac{2kT_{\rm in}}{m_{\rm p} g_{\rm eff}}.
\end{equation} 

Strictly speaking, the adopted dissipation at the bottom of the layer is a very simplified picture, as the energy dissipation is in general distributed along the vertical coordinate in a way that is dependent on the unknown mechanisms involved. 
As long as diffusion approximation is valid, vertical distribution of the dissipation processes leads to two systematic effects: a decrease in the effective optical depth, and deviation from the effective vertical polytrope used in this section. 
Both are likely to introduce correction factors of the order unity, which  would be easy to calculate once our model is extended to three dimensions with a reasonable set of boundary conditions on the surface of the NS. 

\subsection{Momentum equations}\label{sec:euler}

We start with Euler equations with additional source and sink terms related to the momentum of the matter being accreted and to the friction between the SL
and NS surface. 
Their general vector form is
\begin{equation}\label{E:euler:vector}
\pardir{t}{\vector{v}}+(\vector{v}\cdot \nabla) \vector{v} = -\frac{1}{\rho}
\nabla p +\vector{g} + \mbox{source and sink terms},
\end{equation}
where $\vector{g}$ is gravity without the contribution of centrifugal force and is assumed to be directed along the radius vector. 
At the same time, the surface of the NS is close to being equipotential and thus is deformed due to rotation (we neglect all
the other sources of deformation, such as magnetic fields and non-equilibrium
stresses in the crust), which makes the polar-angle component $g_\theta = -
\frac{1}{r}\pardir{\theta}{\Phi}$, where $\Phi$ is gravitational potential,
non-zero even after vertical integration. 

The radial component of the momentum equation reduces to the hydrostatic equation
considered in Sect.~\ref{sec:vert}.
The two tangential components of the equation are convenient to re-write in
terms of the two scalar quantities normally used in shallow-water
approximation: vorticity,
\begin{equation}\label{E:euler:vort}
\omega = \left[\nabla \times \vector{v}\right]_r = \frac{1}{r\sin\theta}
\left( \ppardir{\theta}{v_\varphi \sin\theta} -
\pardir{\varphi}{v_\theta}\right),
\end{equation}
and divergence,
\begin{equation}\label{E:euler:div}
\delta = \left(\nabla \cdot \vector{v} \right) =
\frac{1}{r\sin\theta}\left( \ppardir{\theta}{v_\theta \sin\theta } + \pardir{\varphi}{v_\varphi}\right).
\end{equation}
Multiplying Eq.~(\ref{E:euler:vector}) by $\rho$ and integrating over the total vertical extent of the SL yields
\begin{equation}\label{E:euler:vint}
\pardir{t}{\vector{v}}+(\vector{v}\cdot \nabla) \vector{v} = -\frac{1}{\Sigma}
\nabla \Pi +\vector{g} + \mbox{source and sink terms}.
\end{equation}
A detailed derivation of the equations for vorticity and divergence is given in Appendix~\ref{sec:derive}.

Taking the radial curl component of Eq.~(\ref{E:euler:vint}) results in
an equation for vorticity
\begin{eqnarray}\label{E:euler:omega}
  \displaystyle  \pardir{t}{\omega} & +& \nabla \cdot (\omega \vector{v})  = 
  -\nabla\times \frac{\nabla \Pi}{\Sigma} 
  +\left( \omega_{\rm d}-\omega\right) \frac{S^+}{\Sigma} \nonumber \\
  \displaystyle   &+&
\left[ (\vector{v}_{\rm d}-\vector{v})\times \nabla \frac{S^+}{\Sigma}\right]_r
  + \frac{1}{t_{\rm fric}} \left( 2\Omega_* \cos\alpha
  -\omega\right),
\end{eqnarray}
where $\Omega_*$ is the angular frequency of the NS. 
The velocity field of the accreting matter $\vector{v}_{\rm d}$ is assumed to be uniform rotation with the angular frequency $c_{\rm K}\Omega_{\rm K}$, where $c_{\rm K}$ is the deviation from the Keplerian rotation law in the disc, set to $0.9$ in all the simulations. 
Vorticity of this velocity field is $\omega_{\rm d} = 2 c_{\rm K}\Omega_{\rm K}\cos \alpha$, where $\alpha$ is
given by Eq.~(\ref{E:mass:sina}).
The last term in~(\ref{E:euler:omega}) describes viscous coupling between the SL and the surface of the NS. 
Our lack of knowledge about the nature and strength of this
coupling is included in the unknown friction timescale $t_{\rm fric}$. 
Preceding terms containing $S^{+}$ appear due to accretion of matter with a given vorticity. 
In addition to this, the right-hand side of the equation includes a baroclinic term $\displaystyle \nabla\times \frac{\nabla \Pi}{\Sigma}$ equal to zero if a fixed equation of state is adopted, or if the distributions of pressure and density are exactly axisymmetric. 
This term can create vorticity through entropy variations. 

Taking divergence of Eq.~(\ref{E:euler:vint}) provides an equation for $\delta$
\begin{equation}\label{E:euler:delta}
\begin{array}{l}
 \displaystyle \pardir{t}{\delta} = \left[ \nabla \times (\omega \vector{v})\right]_r
- \nabla^2 \left(\frac{v^2}{2}+\Delta\Phi\right) - \nabla \cdot \left( \frac{1}{\Sigma}
\nabla \Pi \right)\\
 \displaystyle \qquad{} -\delta \frac{S^+}{\Sigma} + \left( \vector{v}_{\rm d} - \vector{v}\right)
\cdot \nabla \frac{S^+}{\Sigma}
- \frac{\delta}{t_{\rm fric}}.\\
\end{array}
\end{equation}
The term $\Delta\Phi = \frac{1}{2}\Omega_*^2R_*^2$ originates from the
rotational deformation of the NS. 

\subsection{Energy conservation}\label{sec:energy}

In general form, energy conservation implies \citep{SP06}:
\begin{equation}\label{E:energy:general}
\ppardir{t}{\frac{1}{2}\rho v^2 + \varepsilon}+\nabla \cdot \left(
\left[\frac{1}{2}\rho v^2 + \varepsilon + p \right]
\vector{v}\right) = q_{\rm NS} + q^+ - q^{-},
\end{equation}
where the right-hand side accounts for heat exchange with the NS ($q_{\rm NS}$), heat released within the layer ($q^+$), and radiation losses $q^{-}$. After integration, all the $q$ quantities result in corresponding capital $Q$ quantities: fluxes through the surface and energy release per unit area.

We treat the hydrodynamics of the SL as ideal, though the numerical solution
techniques used (described later in Appendix~\ref{sec:numerics}) provide
dissipation on small scales close to the spatial resolution used.
If momentum transfer is dominated by turbulent motions forming a direct
cascade similar to Kolmogorov cascade \citep{turbulence} where energy is
transferred from larger to smaller scales, the exact nature and properties of
the viscous dissipation at the small scales are irrelevant. However, conservation of energy implies that viscous dissipation should act as an additional source of internal energy. All the kinetic energy lost by the flow should reappear as heat.
We assume this heat to appear at the bottom of the flow and to
diffuse upwards leaving the SL from its upper surface.
As already mentioned in Sect.~\ref{sec:vert}, for dissipation taking place throughout the volume, this introduces a systematic uncertainty of the order unity in the equations of vertical structure.

Taking into account momentum conservation, friction, and viscous dissipation,
and integrating the energy equation vertically, we end up with the equation
\begin{equation}\label{E:energy:Eint}
\displaystyle \pardir{t}{E} + \nabla \cdot \left( E \vector{v}\right) =-\delta
\Pi + Q^+ - Q^- + Q_{\rm NS} + Q_{\rm acc},
\end{equation}
where $Q^+$ is the heat released in the spreading layer, $Q_{\rm NS}$ is the
heat received from the neutron star, $Q^-$ is the radiation flux lost from
the surface, and the additional term $Q_{\rm acc}$ corresponds to the thermal
energy introduced with the accreting matter and released during its mixing
with the pre-existing material.
Vertically integrated pressure and energy are related by~(\ref{E:vert:Eint}). 
Dissipation is calculated as kinetic energy lost by the flow:
\begin{equation}\label{E:energy:qplus}
\displaystyle Q^+ = -\Sigma \vector{v} \cdot \left.\frac{\diff\vector{v}}{\diff t}\right|_{\rm dissipation},
\end{equation}
where the dissipation in velocity is related both to the friction term in Eqs.~(\ref{E:euler:omega}) and (\ref{E:euler:delta}) and with numerical dissipation on the grid scale, which we discuss in detail in Appendix~\ref{sec:numerics}. The energy radiated away from the surface is set by radiation energy diffusion (see Sect.~\ref{sec:vert}):
\begin{equation}\label{E:energy:qminus}
\displaystyle Q^- = \frac{cg_{\rm eff}}{\varkappa }\left(1-\beta\right),
\end{equation}
where $\varkappa$ is the Rosseland average opacity which we assume to be equal to the
Thomson scattering opacity, $\varkappa = \varkappa_{\mathrm{T}} \simeq 0.34{\rm \, cm^2\,g^{-1}}$. 
If $\beta$ approaches zero, the SL becomes a `levitating layer' supported
mainly by radiation pressure \citep{IS99}. For $\beta \ll 1$, the energy loss
term is nearly independent of the physical conditions inside the layer.

Additional terms related to the accreting matter are more difficult to
constrain from a physical point of view. It is natural to assume that the
initial temperature of the newly introduced material is non-zero, and hence
the increase in surface density is accompanied with an increase in surface energy density as well.
In addition, as the velocities of the accreting and the pre-existing matter are in general different, some of the kinetic energy is dissipated during the process of mixing (the exact physical mechanism could be kinetic, hydrodynamic, or MHD).
Energy and momentum conservation laws predict that the amount of dissipated energy per unit of accreted mass is $(\vector{v}_{\rm d}-\vector{v})^2/2$, hence
\begin{equation}\label{E:energy:qacc}
Q_{\rm acc} = \left( \left(\frac{E}{\Sigma}\right)_{\rm d}  + \frac{1}{2}(\vector{v}_{\rm d}-\vector{v})^2 \right) S^+,
\end{equation}
where again the `d' index corresponds to the properties of the mass source. 

Our approach to the vertical structure, as well as to momentum and energy conservation, is mostly in line with the works of \citet{IS99}, \citet{SP06}, and \citet{IS10}. 
However, momentum and energy equations introduced below assume neither axisymmetry nor stationarity, both of which were crucial for the quasi-stationary, one-dimensional SL model. 
Relaxing these assumptions requires that we specify certain physical mechanisms. 
In particular, the stationary nature of the flow allows us to relate surface density and latitudinal velocity in algebraic form independently of the mechanism of angular momentum loss.  
In addition, \citet{IS99} do not consider rotational deformation of the NS, meaning that a co-rotating atmosphere in their assumptions should strongly concentrate near the equator. 
However, we take into account small equilibrium deformation of the star. Apart from making the simulations more realistic, this equilibrium deformation allows a simple set of initial conditions with uniform surface density and pressure. 

\subsection{Initial conditions}\label{sec:model:init}

The simplest possible initial conditions are constant surface density and pressure in combination with rigid-body rotation.
This may be achieved if the rotation rate is exactly equal to the rotation frequency of the NS, $\Omega_*$, and the deformation of the NS makes its surface equipotential.
Configuration is stable and may survive for simulation times vastly exceeding the durations of the runs used in this work. 
In Sect.~\ref{sec:tests:ND}, we use this initial condition configuration to check the numerical stability and dissipation of our numerical scheme. 

Vorticity of a rigid-body rotation is
\begin{equation}\label{E:IC:vort}
  \omega_{\rm init} = 2\Omega_* \cos\theta.
\end{equation}
As the motions are limited to pure rotation, initial divergence is strictly zero. 
To the basic initial condition set, a small (5\%) perturbation was added in the form of an over- or under-density. 
The perturbation is designed as an entropy variation not affecting the pressure distribution. 

\begin{figure*}
\includegraphics[width=\textwidth]{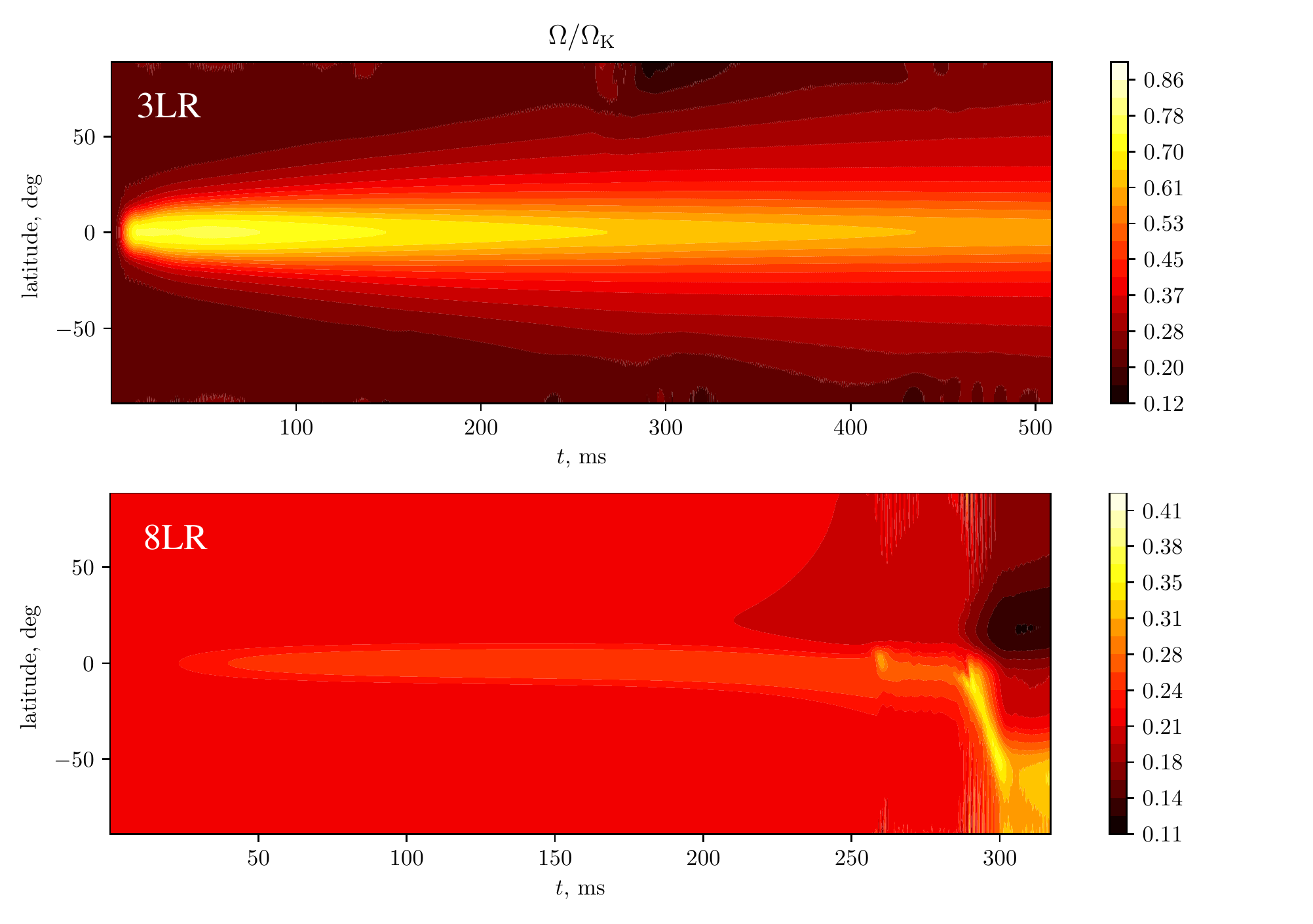}
\caption{Time-latitude diagrams for longitudinally averaged angular frequency normalised to the Keplerian rotation frequency at the radius of the star $R_*$. The upper and lower panels correspond to the rapid accretion model {\tt 3LR} and {\tt 8LR}, respectively.}
\label{fig:twoomegas}
\end{figure*}

\begin{table*}\centering
\caption{Spreading layer simulations. }
\label{tab:mods}
\bigskip
\begin{tabular}{lccccccc}
\hline
Model ID & dimensions & $t_{\rm max}$  & $\Sigma_{\rm init}$ & $\dot{M}$ & PDS\tablefootmark{a} & comments\\
&& s & \gcm & $\Msun\ \mbox{yr}^{-1}$ & s & &\\
\hline
3LR & 128$\times$256 & 0.5 & $10^8$ &   $10^{-3}$   &   0.35--0.5 & \\
3HR & 256$\times$512 & 0.08 &  $10^8$ &  $10^{-3}$   &   & & \\
3LRinc & 128$\times$256 & 0.6 & $10^8$ &   $10^{-3}$    &  0.4--0.6 &  $i=\uppi/4$\\
8LR & 128$\times$256 & 0.32 & $10^4$ &   $10^{-8}$    &  0.27--0.32 & \\
8HR & 256$\times$512 & 0.07 &  $10^4$ &   $10^{-8}$    & & \\
3LRoff & 128$\times$256 & 1.0 & $10^8$ &   0   &    0.5--0.68 & starts with the end of 3LR \\
\hline
\end{tabular}
\tablefoot{
\tablefoottext{a}{Integration limits for calculations of the PDS.}
}
\end{table*}

\begin{figure*}
\includegraphics[width=\textwidth]{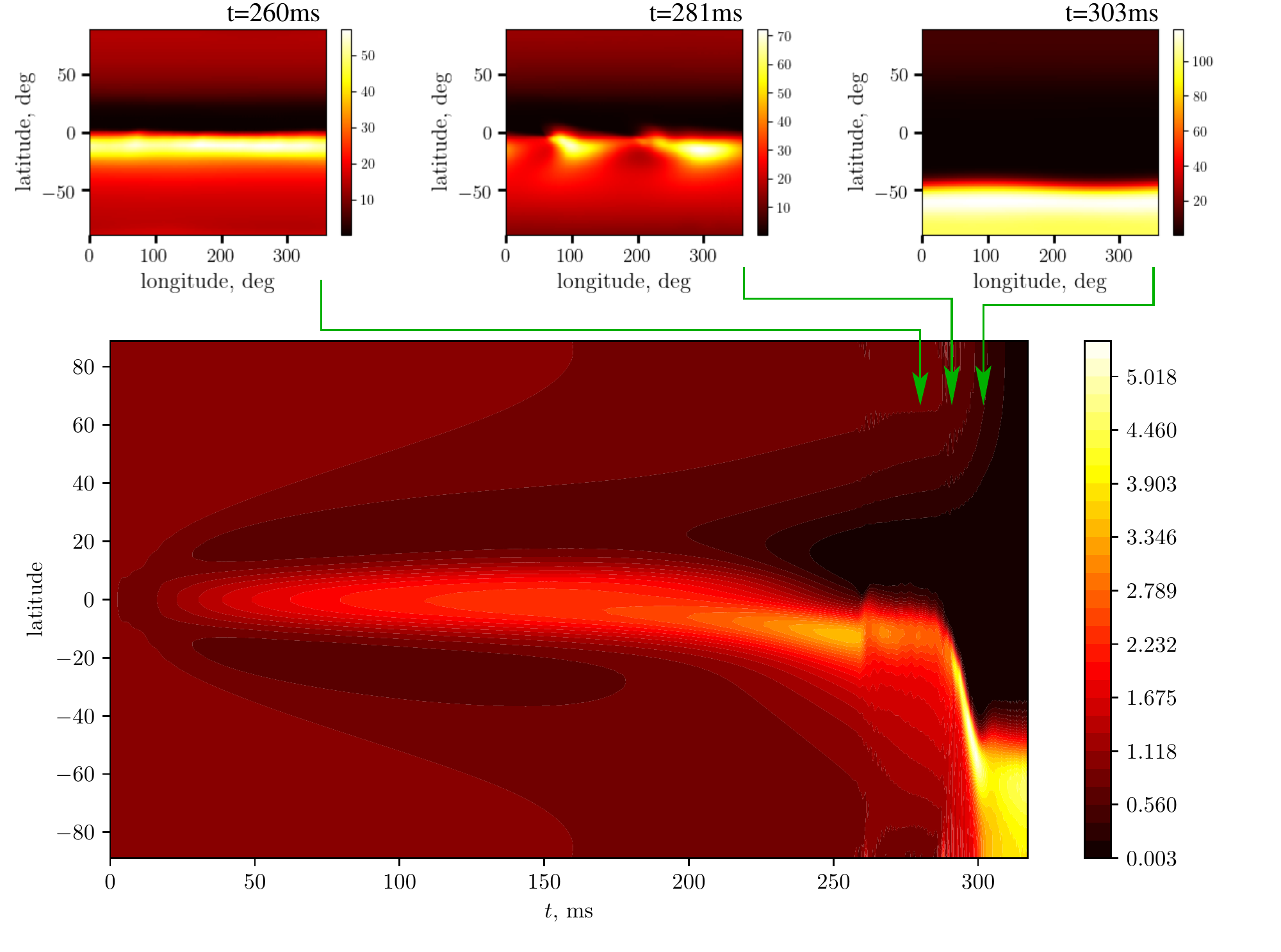}
\caption{Time-latitude diagram for longitudinally averaged surface density (normalised by the surface density averaged over the sphere) in the realistic-accretion-rate {\tt 8LR} simulation. Upper panels show three snapshots of surface density ($10^4\gcm$ units) during the later stages of the heating instability development.} 
\label{fig:symbreak}
\end{figure*}

\begin{figure*}
\includegraphics[width=\textwidth]{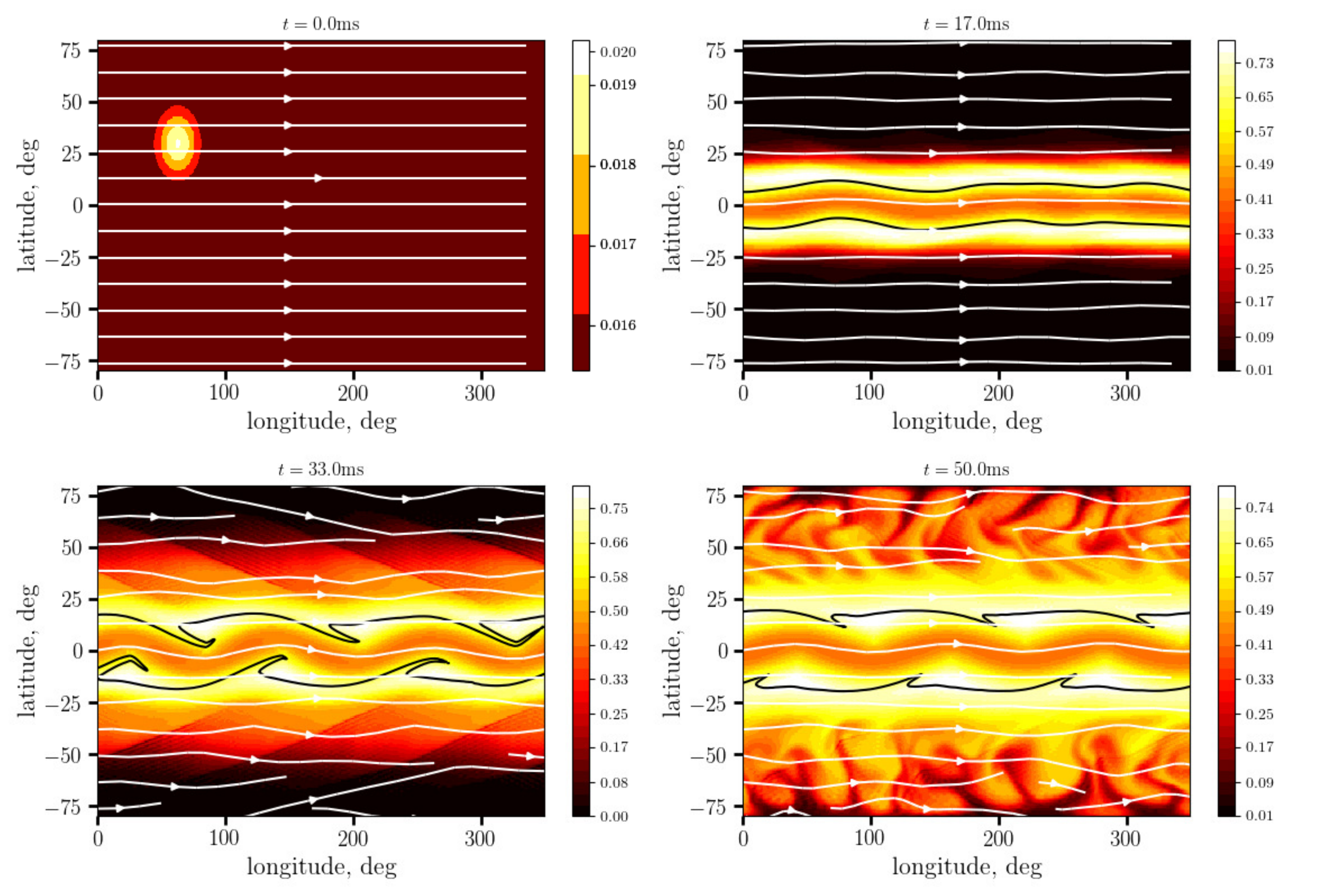}
\caption{Radiation flux (in Eddington units, $GMc/\varkappa R_*^2$) snapshots for the model {\tt 3HR}. White streamlines show the velocity field and the black contour corresponds to the accretion tracer $a=0.5$. For all but the first snapshot, newly accreted matter dominates between the black contours. 
 }\label{fig:HRev}
\end{figure*}

\section{Results}\label{sec:res}

\subsection{Model setup}\label{sec:mods}

The equations derived in Sect.~\ref{sec:model}, including the equations of mass (Eq.~\ref{E:mass:sigma}), momentum (Eqs.~\ref{E:euler:omega} and \ref{E:euler:delta}), and energy (Eq.~\ref{E:energy:Eint}) conservation, were solved using our 2D spectral modelling code.
We refer to Appendix~\ref{sec:numerics} for a detailed description of the numerical techniques. 
There, we also describe the tests for numeric performance, stability, and accuracy. 
We list all the SL models calculated for this paper in Table~\ref{tab:mods}. 
Letters `LR' and `HR' in a simulation ID
always refer to `low' (128$\times$256) or `high' (256$\times$512) resolution. Consistency between the corresponding low- and high-resolution runs is an important test for numerical effects (noise and diffusion). Below, we describe the setups of all these models, while the description of the results is given in the following sections. 

All the models include a NS rotating with a spin period $P_{\rm spin} = 3$\,ms. 
Initially, all the matter on the surface rotates together with the star as a rigid body. 
As we so far do not include any friction with the NS surface, rotation of the star affects the results only through the initial conditions and the deformation of the stellar surface (potential term in Eq.~\ref{E:euler:delta}). 

\begin{figure*}
\includegraphics[width=0.9\textwidth]{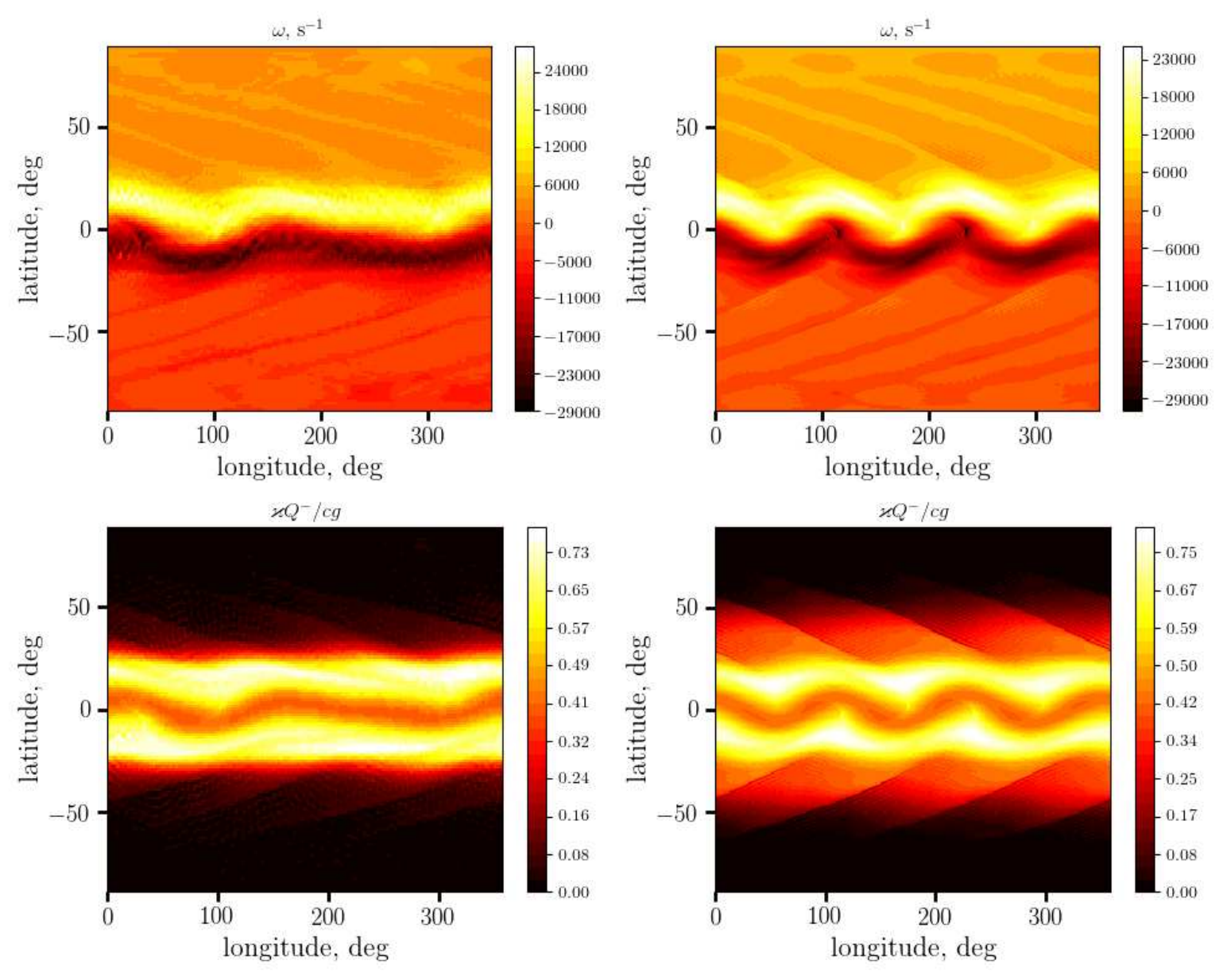}
\caption{Resolution effects after $t=0.03\,$s (ten spin periods) of evolution for the high-accretion-rate models {\tt 3LR} (low numerical resolution; left panels) and {\tt 3HR} (high numerical resolution; right panels). 
Top: Vorticity maps. Bottom: Emitted radiation flux $Q^-$ (normalised by the local Eddington value) map.
 }\label{fig:reseffects}       
\end{figure*}

Most models include a mass source corresponding to a steady-state accretion from a thin disc. 
To avoid strong velocity gradients causing high-frequency noise, the mass accretion rate approaches the steady-state value smoothly, following the exponential law
\begin{equation}
    \dot{M}(t) = \dot{M}_0 \left( 1-e^{-t/t_{\rm turn-on}}\right),
\end{equation}
where the turn-on timescale was set to $10P_{\rm spin}$ for all the simulations. 
The spatial distribution of the source of mass is always a Gaussian function of $\cos\alpha$ as given by Eq.~(\ref{E:mass:source}) with the standard deviation of $\cos \alpha_0 =0.1$, corresponding to the width of about $6\degr$.
Rotation rate of the newly accreted material is set to $c_{\rm K} =0.9$ in  Keplerian units. 

As the timescales observed in real LMXBs differ by six orders of magnitude, it is difficult to perform a single realistic simulation resolving the dynamic-timescale variability over a timescale that is sufficient to see the changes in the SL structure. We use two approaches to avoid this difficulty: first, we consider  `enhanced accretion' with $\dot{M} = 10^{-3}\Msun\ \mbox{yr}^{-1}\simeq 6\times 10^{20}\ \mbox{g\,s}^{-1}$ (models {\tt 3LR} and {\tt 3HR}); secondly, we consider accretion on top of a thin layer with $\Sigma_0 = 10^{4}\gcm$. Both tricks shorten the effective evolution timescales by several orders of magnitude. We expect the hydrodynamics to be equally effective in both configurations even though the radiation timescales are vastly different. 

As an alternative to both these approaches, we calculate a model reproducing the evolution of a spreading layer after switching off the mass source (model {\tt 3LRoff}). It starts with the final snapshot of {\tt 3LR} and then gradually cools down for another 0.5\,s. 
Finally, in the model {\tt 3LRinc}, we consider the case where the source is inclined with respect to the initial rotation plane.

\subsection{General properties}\label{sec:res:gene}

Each simulation covers several tenths of a second, which is significantly longer than the dynamical timescales, and for the parameters
of our simulations is comparable to the timescale on which the initial mass content of the layer is replaced by newly accreted matter. 
Accretion is concentrated (everywhere except {\tt 3LRinc}) at low latitudes. 
As the surface mass and energy density increase near the equator, matter is pushed toward the poles. 
Surface density in the equatorial region tends to become larger due to accumulation of the accreted material, but often becomes lower, as the mixing between the streams moving at different velocities leads to high levels of dissipation (see  Eq.~\ref{E:energy:qacc}). 
Equatorial regions tend to spin faster in all the models (see Fig.~\ref{fig:twoomegas}), but the excess angular momentum is redistributed by  oblique waves (for rapid accretion modes) or by the heating instability (for {\tt 8LH/8HR}).
The main difference between the models with different mass accretion rate is the role of cooling: for rapid accretion, heating occurs much faster and the layer effectively accumulates heat, while for the models {\tt 8LR/8HR}, radiation efficiently cools down some portions of the flow.
  
For the low mass accretion rate (model {\tt 8LR}), the evolution of the SL is illustrated by Fig.~\ref{fig:symbreak}.
We note the decrease in surface density at latitudes of about $\pm20\degr$ ($t\simeq$50--100ms) due to heating, the gradual north-south symmetry breaking between $t\simeq$200 and 250\,ms, and later dynamical-timescale evolution. In these simulation runs, $\beta \simeq 0.99\div 1$, close to $1$ with an accuracy of about $10^{-5}$ outside the regions of intense mixing and heating instability. 

For the  high mass accretion rate models, {\tt 3LR} and {\tt 3HR}, the main process driving the subsequent evolution is the velocity contrast between the new and old matter, leading to a shear instability. 
A sequence of snapshots of the radiation flux and velocity field for model {\tt 3HR} is shown in Fig.~\ref{fig:HRev}.  
The gas-to-total pressure ratio $\beta$ for enhanced accretion models varies between about $0.1$ in the accretion region and $\sim 0.9$ near the poles. 
 
As we do not include the sinks in this paper, no quasi-stationary picture is expected to be reached. 
However, heating and velocity gradients created by accretion lead to at least two important dynamical effects relevant for the dynamics of SLs. In the two groups of simulations, with `enhanced' and `normal' mass accretion rates, two different instabilities emerge. For the realistically low mass accretion rate ($\dot{M} = 10^{-8}\Msun\ \mbox{yr}^{-1}$) and low initial surface density ($\Sigma_0 = 10^4\gcm$), the equatorial belt forming out of the accreting matter during the first milliseconds of accretion is cooled efficiently, and most of the subsequent dissipation takes place at higher latitudes where the newly added material mixes with the old, slowly rotating NS atmosphere. 
This results in a {\it heating instability}: local displacement of the cool equatorial belt material leads to increased dissipation on the opposite side of the equator, which results in a pressure gradient increasing the initial displacement. 
This is easily seen in Fig.~\ref{fig:symbreak}, where the later-time evolution (starting at approximately $t\sim 0.2$\,s) is marked by a gradual and then dynamical-timescale development of a strong mass asymmetry between the southern and the northern hemispheres.  Development of the instability also breaks the axial symmetry, especially during the period of rapid evolution. As a large amount of matter migrates between the polar and equatorial regions, some part of the flow acquires very large, and some very slow rotational frequency (even smaller than the rotation velocity of the NS).

At the same time, for the case of enhanced accretion, cooling and heating timescales are much longer, and all the observed dynamical effects are purely hydrodynamical. They are reasonably reproduced by the low-resolution simulations, though increasing resolution reveals more details and adds regularity to the observed turbulent patterns (see Fig.~\ref{fig:reseffects}). Spectral simulations are able to capture the oblique wave patterns and `curling' of the equatorial spreading layer belt even with the low numerical resolution runs. For the high-resolution runs, more fine-scale substructure starts to appear.

\subsection{Angular momentum transport within the layer}

\begin{figure}
\includegraphics[width=1.0\columnwidth]{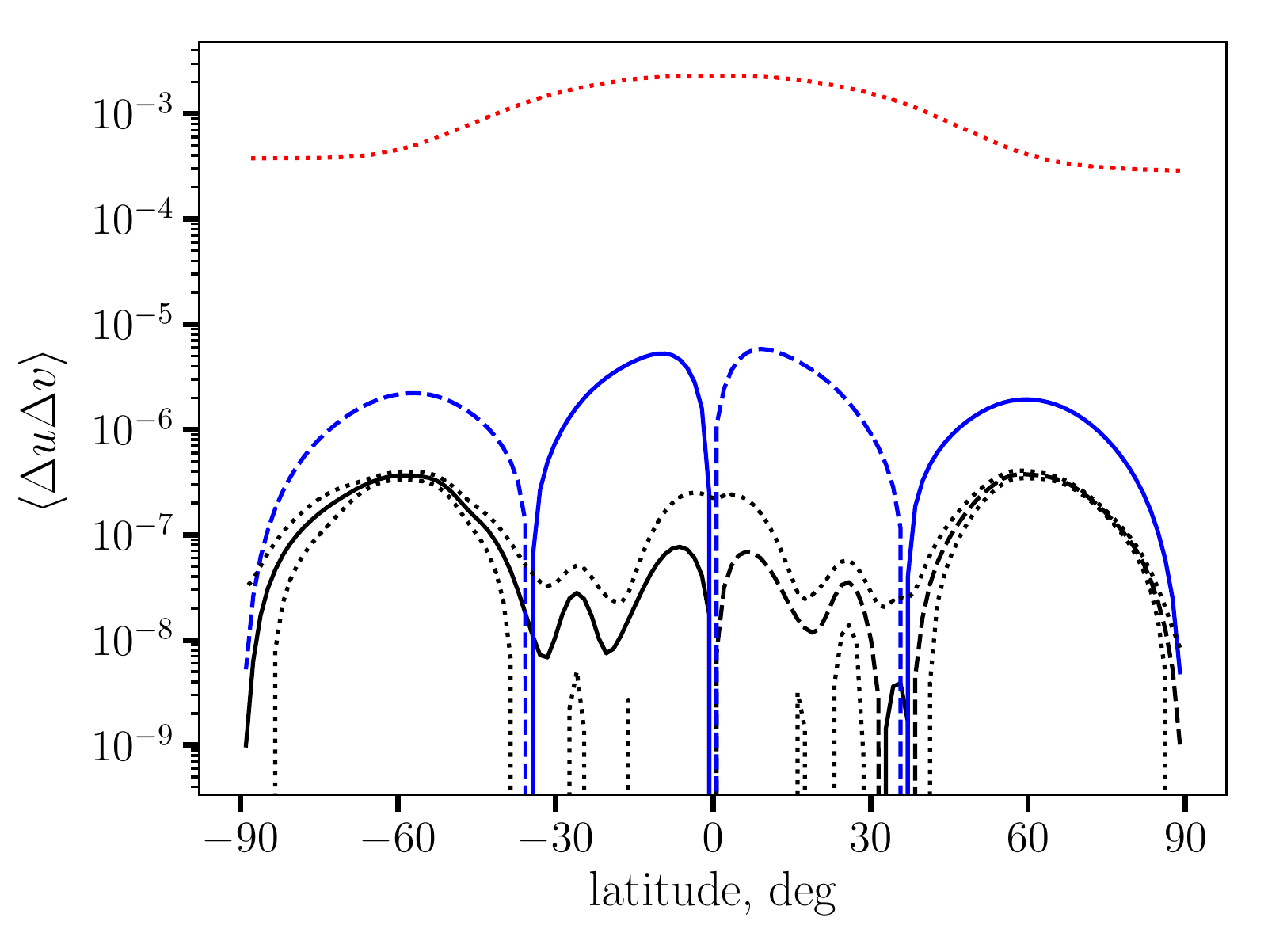}
\caption{Azimuthally- and temporally averaged dynamical quantities for the high-accretion-rate model {\tt 3LR}.
Visualised quantities are Reynolds stress (black), mean velocity product $v_\theta v_\varphi$ (blue), and sound velocity (red dotted) as functions of latitude. All the quantities were averaged over the period of time between 0.1 and 0.3\,s, and over the azimuthal angle. Solid and dashed lines correspond to positive ({southward motion} or momentum transfer) and negative  quantities ({northward}). Dotted black curves show the $1\sigma$ standard deviation interval for the Reynolds stress.  
 }\label{fig:rxy3}       
\end{figure}

The high-accretion-rate models {\tt 3LR} and {\tt 3HR} demonstrate a system of oblique waves generated by the velocity discontinuity. The discontinuity itself in our simulations is a consequence of the initial conditions. In real sources,  the existence of old material co-rotating with the star is a consequence of angular momentum exchange with the star. 
Whether or not the velocity drop exists in a quasi-stationary case with a sink and friction is an open question that may be resolved by running a simulation with sink terms for a  sufficiently long time. 
In the new, rapidly rotating part of the flow, a system of standing waves rapidly evolves into a non-linear regime, and forms a non-axisymmetric wiggle structure seen in Fig.~\ref{fig:reseffects} that is subsequently smeared off. At large latitudes, where the old, slowly rotating matter dominates, the waves create a correlation between orthogonal velocity components. Velocity correlation provides a Reynolds stress component 
\begin{equation}  
  \displaystyle  T_{\theta \varphi} = \left\langle \Sigma (v_\theta - \langle v_\theta\rangle)  (v_\varphi - \langle v_\varphi\rangle )\right\rangle,
\end{equation}
where angular brackets $\langle\ldots \rangle$ denote averaging in time and longitude.
In Fig.~\ref{fig:rxy3}, we show the value of $T_{\theta\varphi}$ calculated for the model {\tt 3LR} for the period of time 0.1--0.3\,s after the start. Apparently, Reynolds stress is small in comparison to pressure but surprisingly stable in its sign, showing a clear poleward angular momentum transfer.
While the SL itself appears to approach a quasi-stationary axisymmetric state on a  timescale of about 1s, higher latitudes still show variability and non-axisymmetry up to the end of the simulation. 
In real astrophysical sources, the mass accretion rate is variable on subsecond timescales, meaning that even if the Reynolds stress is a reaction to the variations in mass accretion rate, it should always be present.

\begin{figure}
\includegraphics[width=1.0\columnwidth]{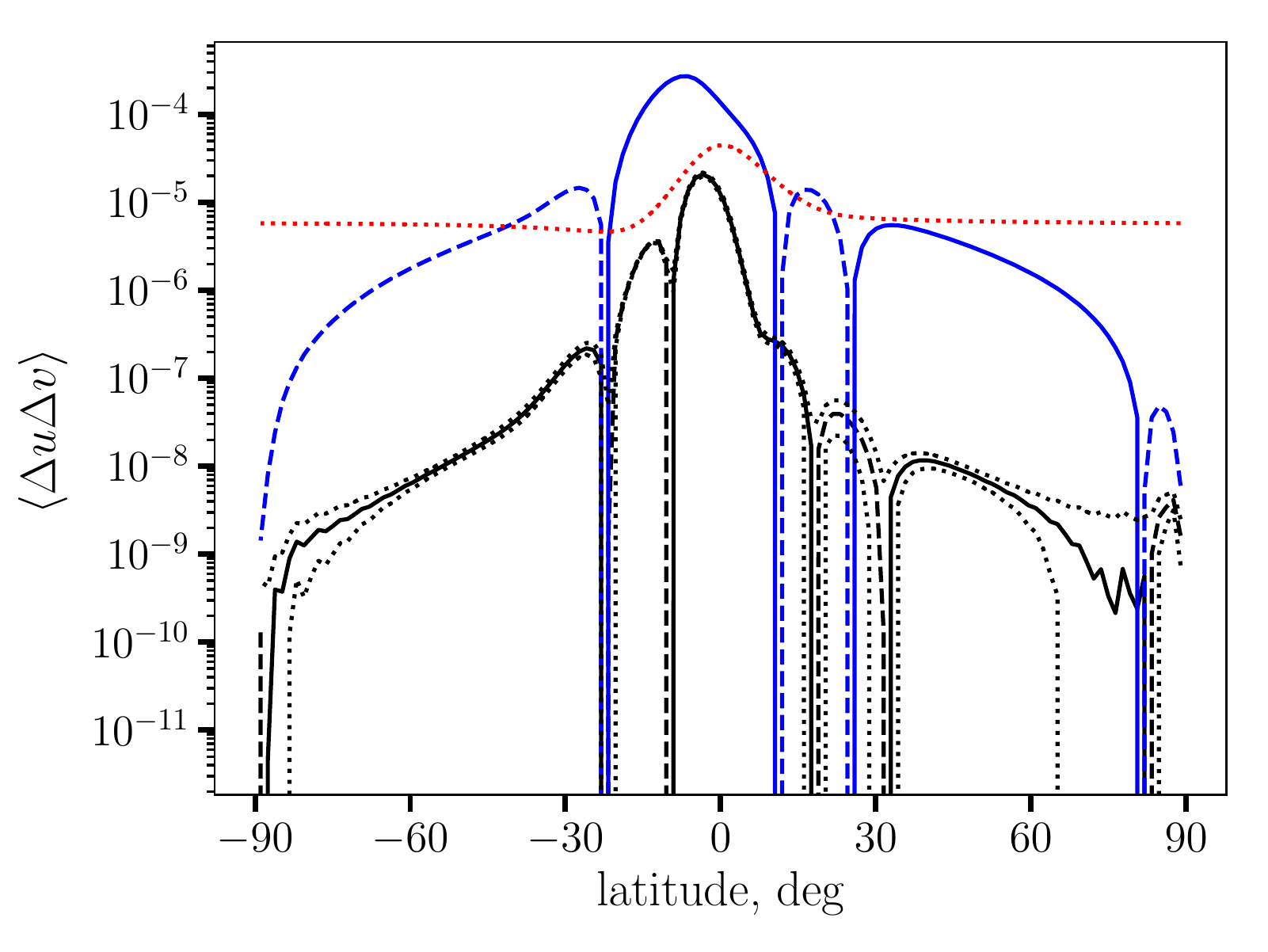}
\caption{Azimuthally- and temporally averaged dynamical quantities for the realistic-accretion-rate model {\tt 8LR}. Symbols and quantities are the same as in Fig.~\ref{fig:rxy3}.
    Averaging was performed from $t=0.2$ to 0.28\,s, when the initial hemisphere asymmetry is developed due to heating instability. 
 }\label{fig:rxy8}       
\end{figure}

The existence of a small but significant hydrodynamical stress  suggests a long local viscous timescale corresponding to oblique-wave-mediated angular momentum transfer. Near the poles, Reynolds stress partially compensates the advective angular momentum flow caused by compression of the pre-existing polar cap material. Near the equator, at the same time, both advection and viscous transport of angular momentum are directed polewards. 

For the realistic-accretion-rate models, {\tt 8LR} and {\tt 8HR}, heating instability creates a strong flow of mass toward one of the poles. 
At the later stage of this process, when most of the accumulated mass flips to one side, axial symmetry is broken, which effectively creates a very large Reynolds stress spreading the angular momentum of the rapidly rotating matter over latitudes (see Fig.~\ref{fig:rxy8}). 
Reynolds stress rapidly removes the angular momentum from the equatorial stream in both directions (we note the sign change near $-10^{\circ}$ latitude). 

\subsection{Artificial light curves}\label{sec:res:lcurves}

To produce artificial light curves, we use a simplified approach that ignores all the relativistic effects. We choose an inclination of the observer $i_{\rm obs}$ and integrate the bolometric flux $Q^-$ emitted from the surface:
\begin{equation}\label{E:lobs}
    L_{\rm obs} = 4 \int_{\alpha_{\rm obs} < \uppi/2} Q^- \cos\alpha_{\rm obs}\ R^2 \sin \theta d\theta\ d\varphi,
\end{equation}
where 
\begin{equation}
    \cos \alpha_{\rm obs} = \cos\theta \cos i_{\rm obs} + \sin \theta \sin i_{\rm obs} \cos \varphi,
\end{equation}
and $\alpha_{\rm obs}$ is the angle at which the surface element is seen by an observer located at the co-latitude of $i_{\rm obs}$ and at the azimuthal angle of $\varphi = 0$ in the spherical coordinate system. 
Such an approach allows us to reproduce the effects of visibility of any moving features on the surface of the star. 
The factor four in Eq.~(\ref{E:lobs}) allows us to interpret $L_{\rm obs}$ as isotropic luminosity, equal to the actual luminosity for an isotropic source.

\begin{figure}
\includegraphics[width=\columnwidth]{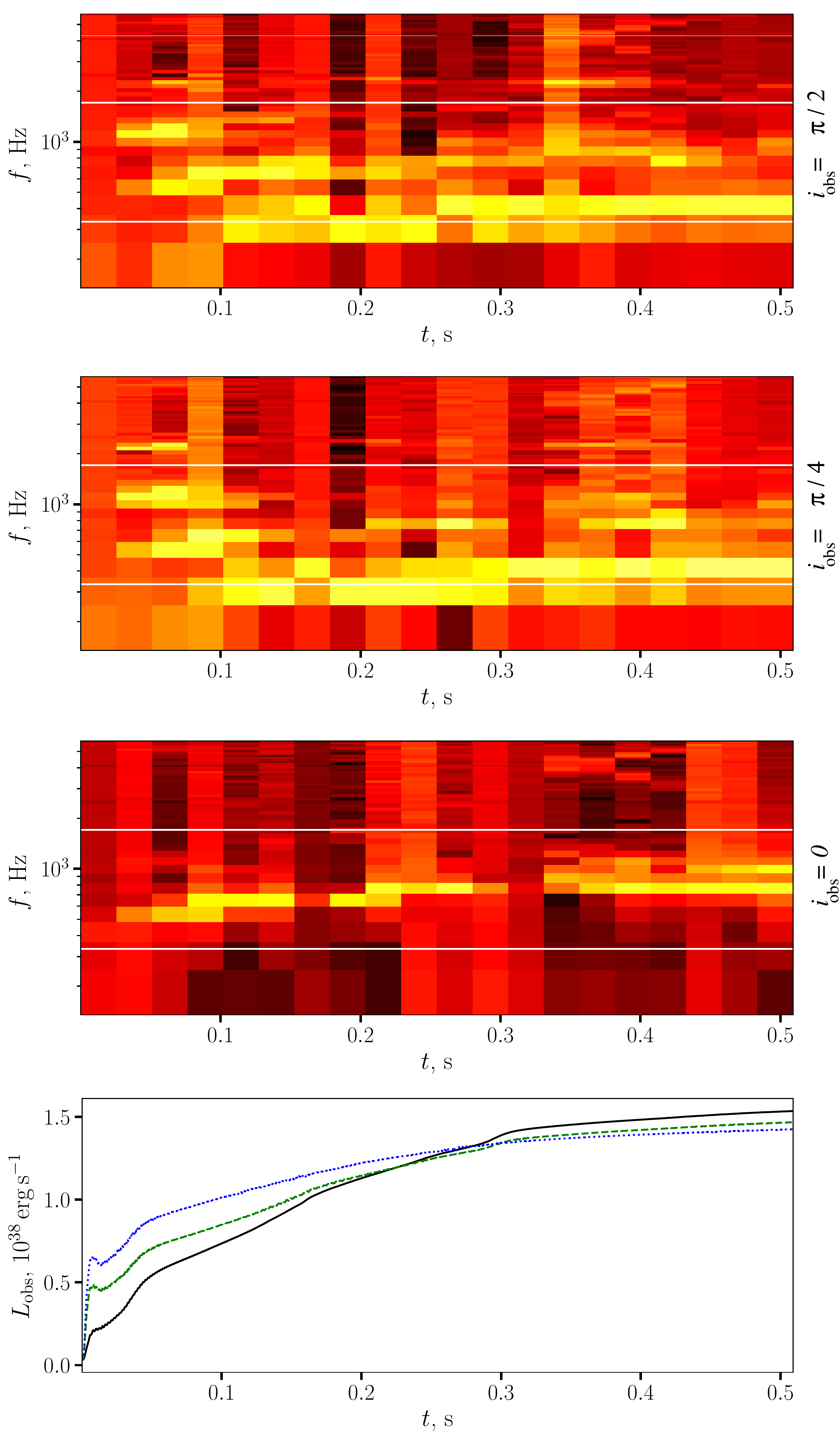}
\caption{Dynamical PDSs ($\log_{10}(f^2 PDS(f) )$ normalised to total power within a single time bin) for the `isotropic luminosity' calculated as described in
  Sect.~\ref{sec:res:lcurves} for the high-accretion-rate model {\tt 3LR}. The three upper panels show dynamical spectra for the observer's
  inclinations of $\uppi/2$, $\uppi/4$, and 0, respectively. White horizontal
  lines show the spin (lower) and Keplerian frequencies. The lowermost panel shows the corresponding light curves: $i_{\rm obs}=\uppi/2$, $\uppi/4$, and $0$ cases are shown with blue dotted, green dashed, and black solid lines.}\label{fig:whistler_3LR}       
\end{figure}

\begin{figure}
\includegraphics[width=\columnwidth]{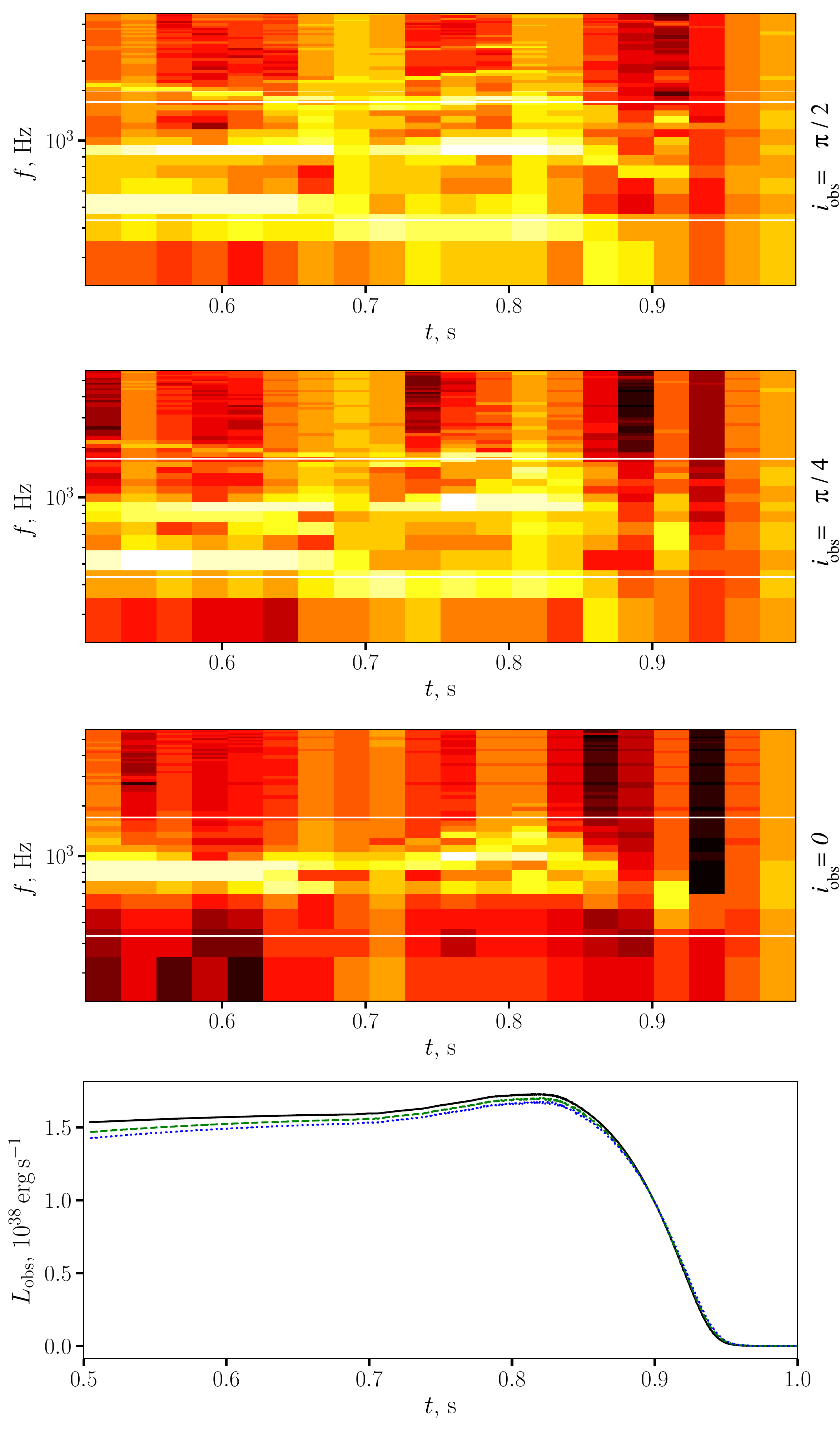}
\caption{Same as Fig.~\ref{fig:whistler_3LR}, but for the model with a turned-off mass source, {\tt 3LRoff}.
 }\label{fig:whistler_3LRoff}       
\end{figure}

\begin{figure}
\includegraphics[width=\columnwidth]{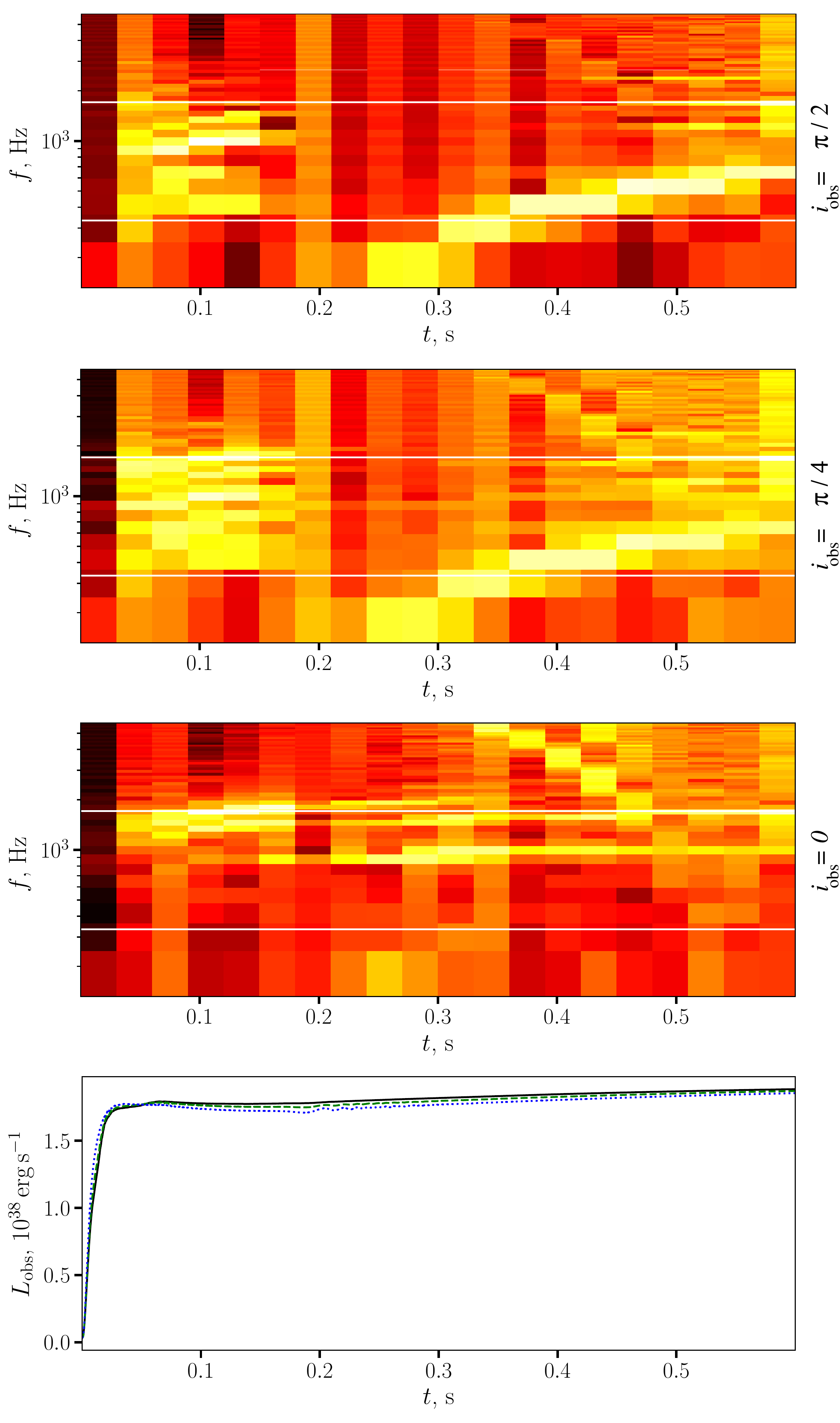}
\caption{Same as Fig.~\ref{fig:whistler_3LR}, but for the model with an inclined source, {\tt 3LRinc}. }\label{fig:whistler_3LRinc}     
\end{figure}

Power density spectra (PDS) were calculated using the standard FFT algorithm \citep{fft} with fractional normalisation. If the light curve is set as a series of observed luminosities, $L_k$, at the equidistant instances of time, $t_k$, the Fourier power density (in Miyamoto normalisation, see e.g. \citealt{miyamoto,nowak99}) is found as a function of frequency, $f$, as
\begin{equation}
    \displaystyle PDS(f) = 2\left| \frac{\sum_k L_k e^{-2\uppi i f t_k}}{\sum_k L_k} \right|^2 \simeq 2\left| \frac{\int L e^{-2\uppi i f t} \diff t}{\int L \diff t} \right|^2.
\end{equation}
The frequency grid on which the PDS is calculated is equally spaced with $\Delta f = 1/T$, where $T$ is the time span. 
Spectral power defined this way is a measure of the relative amplitude of a variability mode. 
For a broad spectral peak, variability amplitude may be estimated as $\sim \sqrt{\sum \Delta PDS }$, where $\Delta PDS$ stands for the excess spectral power associated with the particular spectral detail, and summation is done over the relevant spectral interval. 

\begin{figure*}
\includegraphics[width=0.9\textwidth]{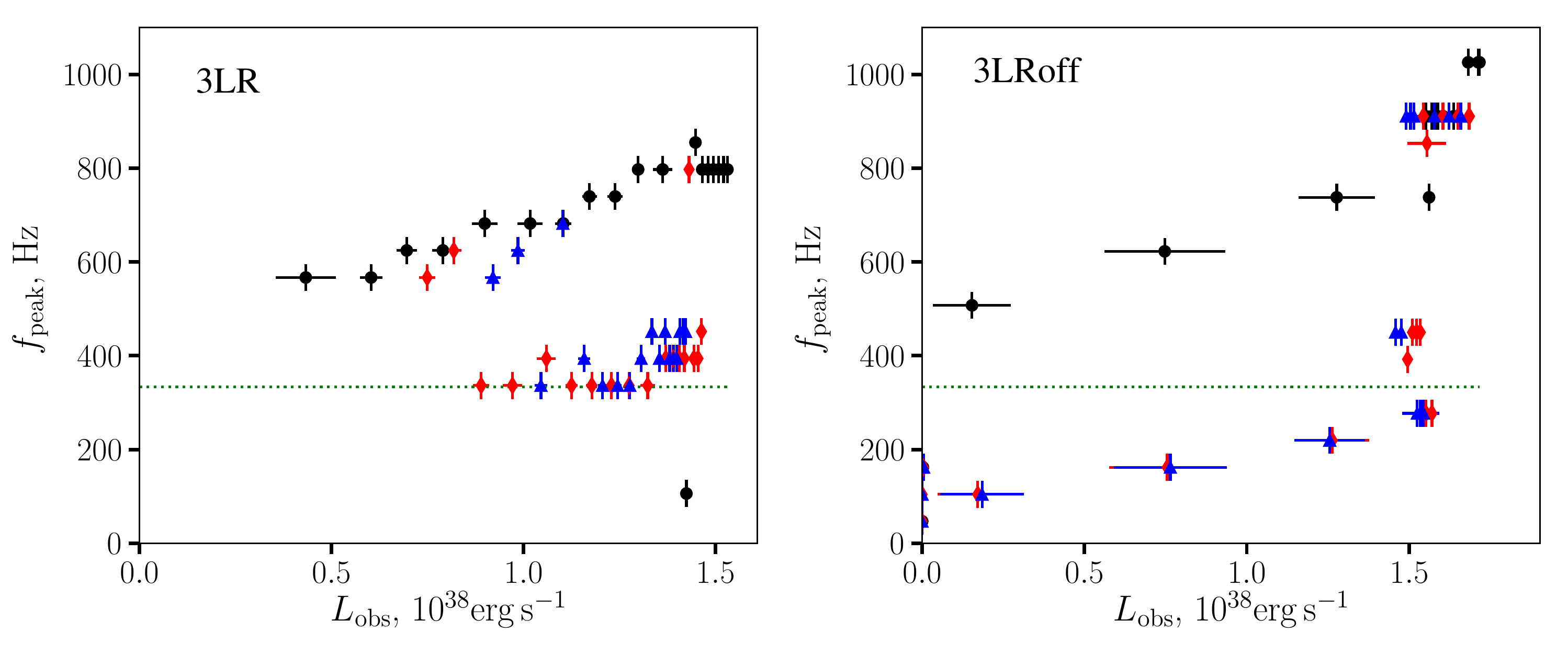}
\caption{Peak frequency (calculated as the position of the maximum of $f \times PDS$) as a function of observed luminosity (calculated using expression~\ref{E:lobs}) for $i_{\rm obs} =0$ (black circles), $\uppi/4$ (red diamonds), and $\uppi/2$ (blue triangles). Simulation runs {\tt 3LR} (left) and {\tt 3LRoff} (right). Error bars show flux dispersion within the time bin and the size of the frequency bin where the maximum was detected. Horizontal dotted green lines show spin frequency. 
}\label{fig:ffreq3}
\end{figure*}

\begin{figure}
\includegraphics[width=0.9\columnwidth]{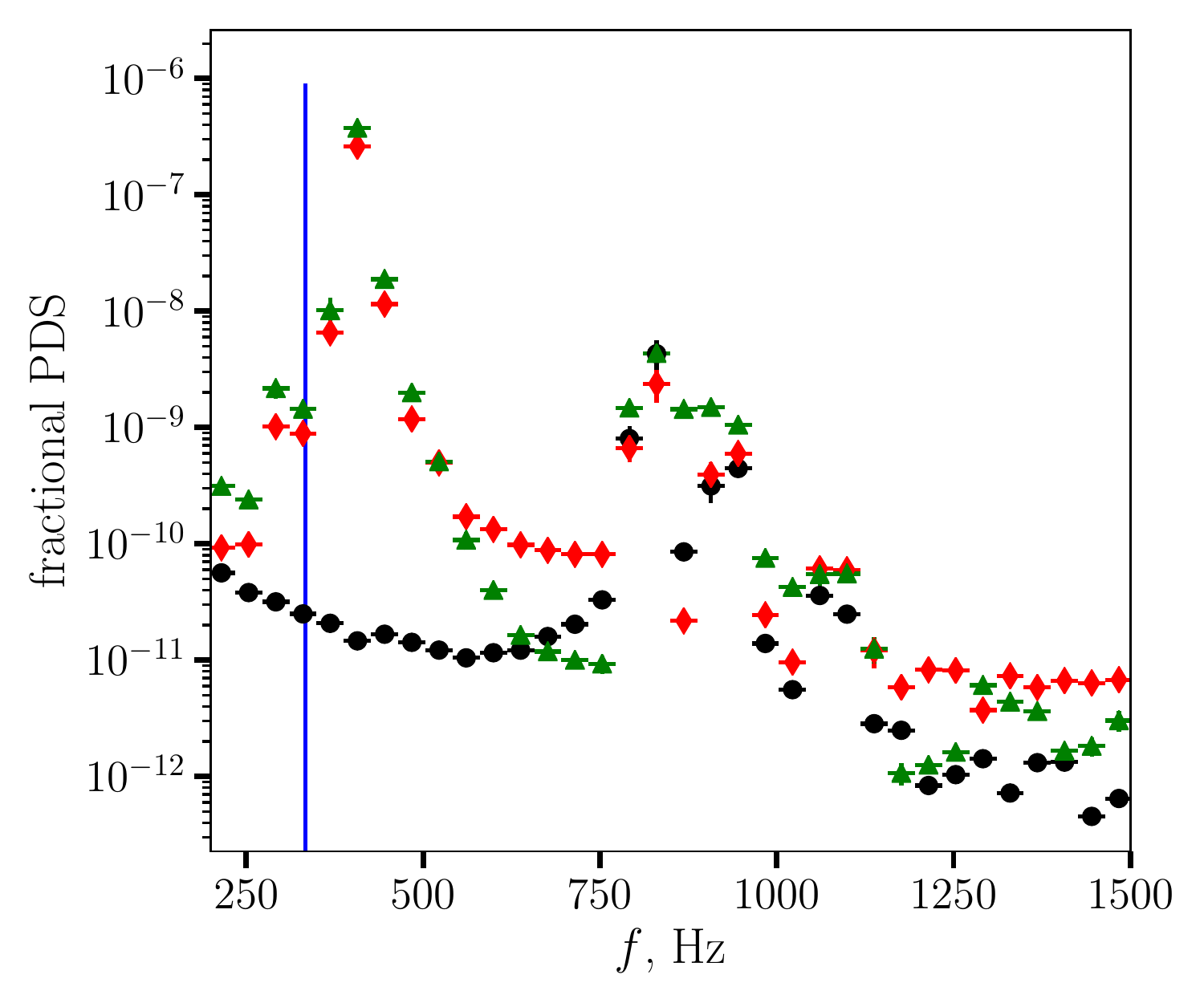}
\caption{Integral PDS (time span 0.2 to 0.5\,s) of the simulation {\tt 3LR} for different inclinations. Black dots correspond to a pole-on observer $i_{\rm obs}=0$, red diamonds to the intermediate inclination of $i_{\rm obs}=\uppi/4$, and green triangles to $i_{\rm obs}=\uppi/2$.
}\label{fig:pds3_3LR}       
\end{figure}

\begin{figure}
\includegraphics[width=0.9\columnwidth]{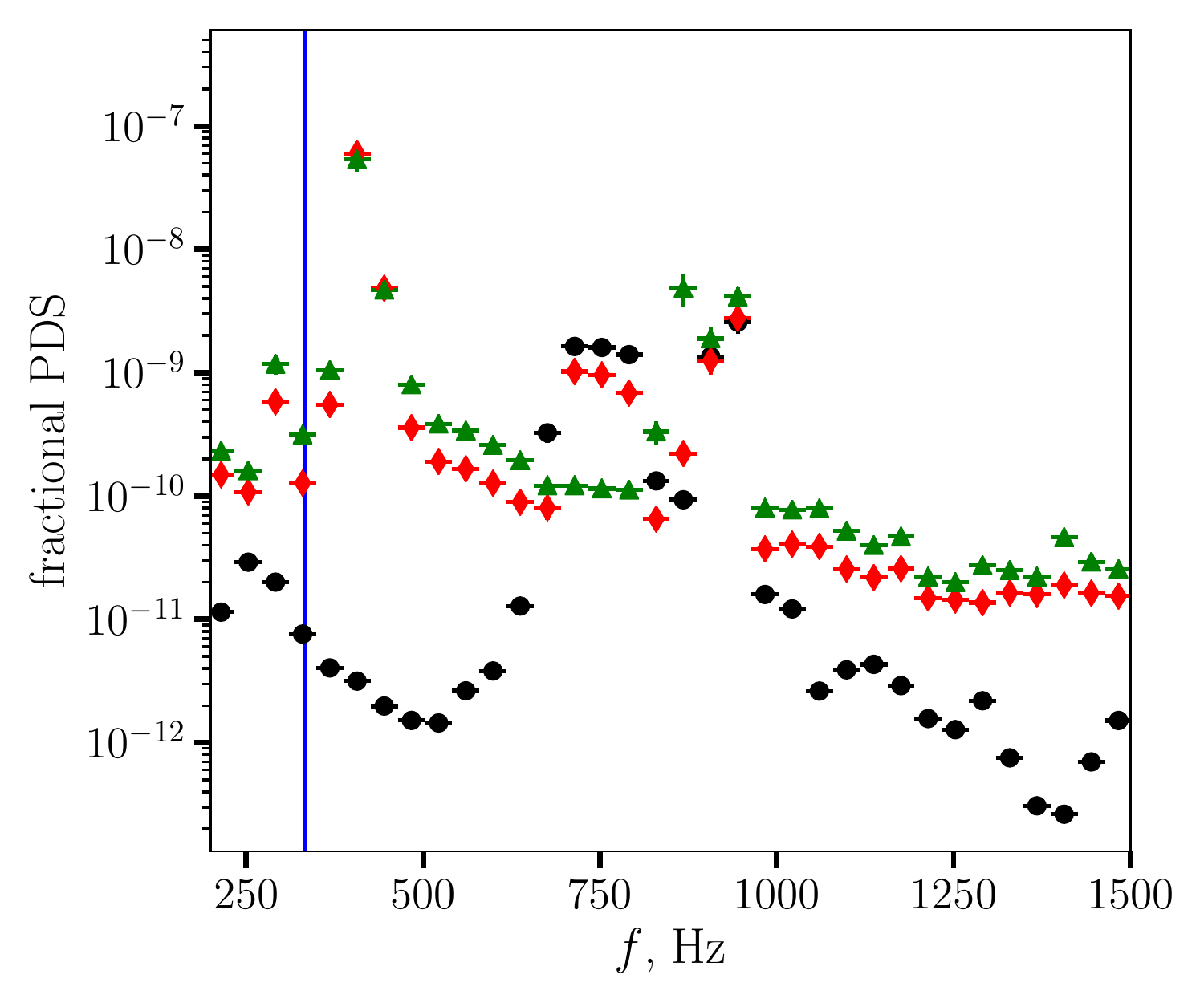}
\caption{Same as Fig.~\ref{fig:pds3_3LR}, but for the simulation {\tt 3LRoff} and between $t=0.58$ and $0.67\,$s.
}\label{fig:pds3_off}       
\end{figure}

In Fig.~\ref{fig:whistler_3LR} we show the dynamic (calculated inside 20 separate time bins) power-density spectra for different $i_{\rm obs}$. We plot the relative PDS multiplied by $f^2$ to decrease the contamination from the low-frequency noise component related to the overall shape of the simulated light curve. Several oscillation modes are visible, one of them for a pole-on observer. Their frequencies evidently correlate with the flux. Most of the non-axisymmetric structures in this simulation are moving slightly faster than the star itself: their contribution is visible just above $\Omega_*$. There is also power at about double spin frequency and in the very beginning near the third harmonic. The perturbation seen during the first $\sim$0.1\,s of the simulation is mostly related to the initial perturbation rotating at the spin frequency. 
However, very little variability is seen by a polar observer, except for a single peak initially close to 1.5$\Omega_*$ and then gradually increasing its frequency towards 700--800\,Hz. This signal is visible for all the inclinations but in general is weaker than the non-axisymmetric modes absent in the pole-on dynamic spectra. 
The properties of this mode fit well into the concept of a latitudinally propagating surface wave moving in a waveguide, similar to the modes considered by \citet{PB04b}. 
We discuss the implications of such an interpretation later in Sect.~\ref{sec:disc}.

In Fig.~\ref{fig:whistler_3LRoff} we show how the dynamic PDS changes when rapid accretion stops (model {\tt 3LRoff}, that starts with the end of simulation {\tt 3LR}). The quasi-periodic features around two-three spin frequencies retain at least for the period when the layer remains hot (before $t\simeq 0.95$\,s). The axisymmetric mode is evidently split into two QPO features. A hint of such a split is visible also at $t\sim$0.3--0.4\,s in Fig.~\ref{fig:whistler_3LR}. All the characteristic features, as in the original simulation with accretion, correlate with flux. 

\begin{figure}
\includegraphics[width=0.9\columnwidth]{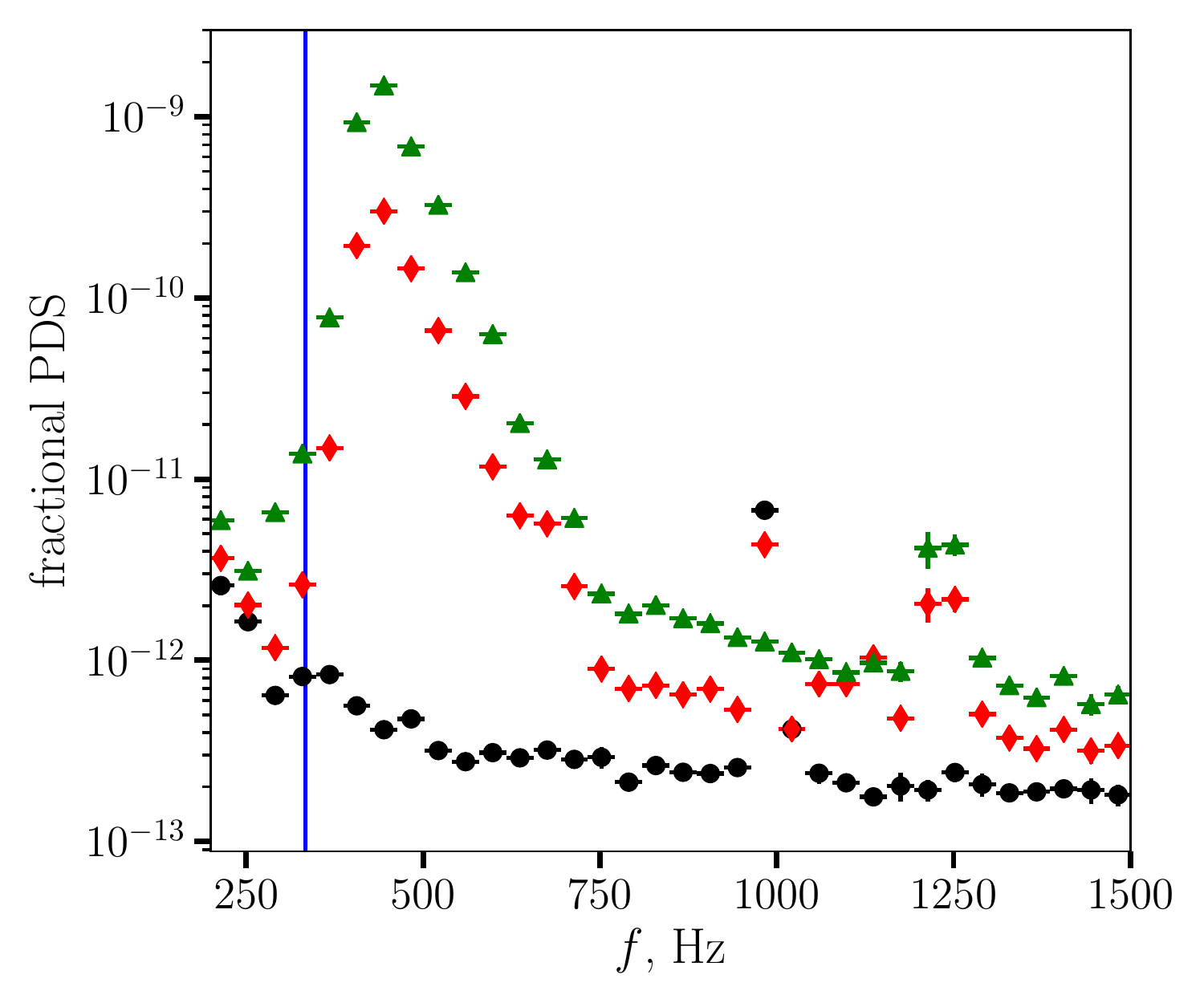}
\caption{Same as Fig.~\ref{fig:pds3_3LR}, but for the simulation {\tt 3LRinc} and for the time span $t=0.35\div 0.5\,$s.  
}\label{fig:pds3_3LRinc}       
\end{figure}

For the inclined simulation, {\tt 3LRinc} (Fig.~\ref{fig:whistler_3LRinc}),
there is an early stage ($0\div 0.2\,$s) of the collision between the two flows inclined to each other. Subsequently, a quasi-axisymmetric configuration forms, and the variability pattern becomes similar to those of the aligned models discussed above. Unlike the aligned case, the observed luminosity changes in very narrow limits, probably because the size of the SL is now determined by the geometry of the inflow rather than by angular momentum transfer. Dissipation is smoothly distributed over the whole latitude range between $-i$ and $i$, and saturates at a level $Q^- \simeq cg_{\rm eff}/\varkappa_{\rm T}$.
The apparent luminosity seen by a pole-on observer is
\begin{eqnarray}\label{E:lobs:inc}
    \displaystyle L_{\rm obs, \, inc} &\simeq& \frac{4 c g_{\rm eff} R_*^2}{\varkappa} \int_{\uppi/2 < \theta < i} \cos\theta \sin \theta \diff \theta \int_{0}^{2\uppi}  \diff \varphi \nonumber \\
     \displaystyle &\simeq& \frac{4\uppi c g_{\rm eff} R_*^2}{\varkappa} \cos^2 i  \simeq 10^{38}\ \mbox{erg\, s}^{-1}. 
\end{eqnarray}
Starting from $t\simeq 0.15\,$s, the pole-on PDS shows one stable peak at about 1\,kHz and a hint of another peak at about 1.5\,kHz, sometimes split in two (see Fig.~\ref{fig:pds3_3LRinc} showing the PDS integrated over the integral 0.4--0.65\,s; a similar picture is seen for $t\sim 0.2\div 0.3$\,s). Unlike the aligned case, non-axisymmetric modes at later stages appear slower than the axisymmetric. This is probably related to the overall change in angular momentum of the layer that is affected by the pre-existing matter rotating in a different direction. 

Peak frequencies in the dynamical PDSs are shown in Fig.~\ref{fig:ffreq3} as functions of flux. There is an evident signal for the pole-on simulated light curve both in the original model with enhanced accretion, and in the switch-off simulation. The axisymmetric mode discussed above dominates for the pole-on case, for which a clear correlation between flux and frequency is observed. For an inclined observer, non-axisymmetric modes are stronger. Interestingly, an inclined source is capable of exciting variability modes with frequencies lower than the spin frequency. 

We also consider PDSs integrated over time intervals where the shapes of dynamic PDSs remain relatively stable and/or show hints of additional spectral features.
In Fig.~\ref{fig:pds3_3LR}, we show such a spectrum for the {\tt 3LR} simulation, computed for $t=$0.2--0.5\,s. 
Pole-on PDS is dominated by a single narrow peak at about 800\,Hz. 
At large inclinations, this single QPO transforms into two, and an additional third peak emerges at about one spin frequency. 
For {\tt 3LRoff}, splitting of the main peak for $i_{\rm obs}=0$ is visible at $t\sim$0.5--0.7\,s (shown in Fig.~\ref{fig:pds3_off}) and later at $t\sim0.8$\,s. Later, the structure of the layer starts rapidly changing due to rapid cooling and switching to the gas-pressure-dominated regime. 

\begin{figure}
\includegraphics[width=0.9\columnwidth]{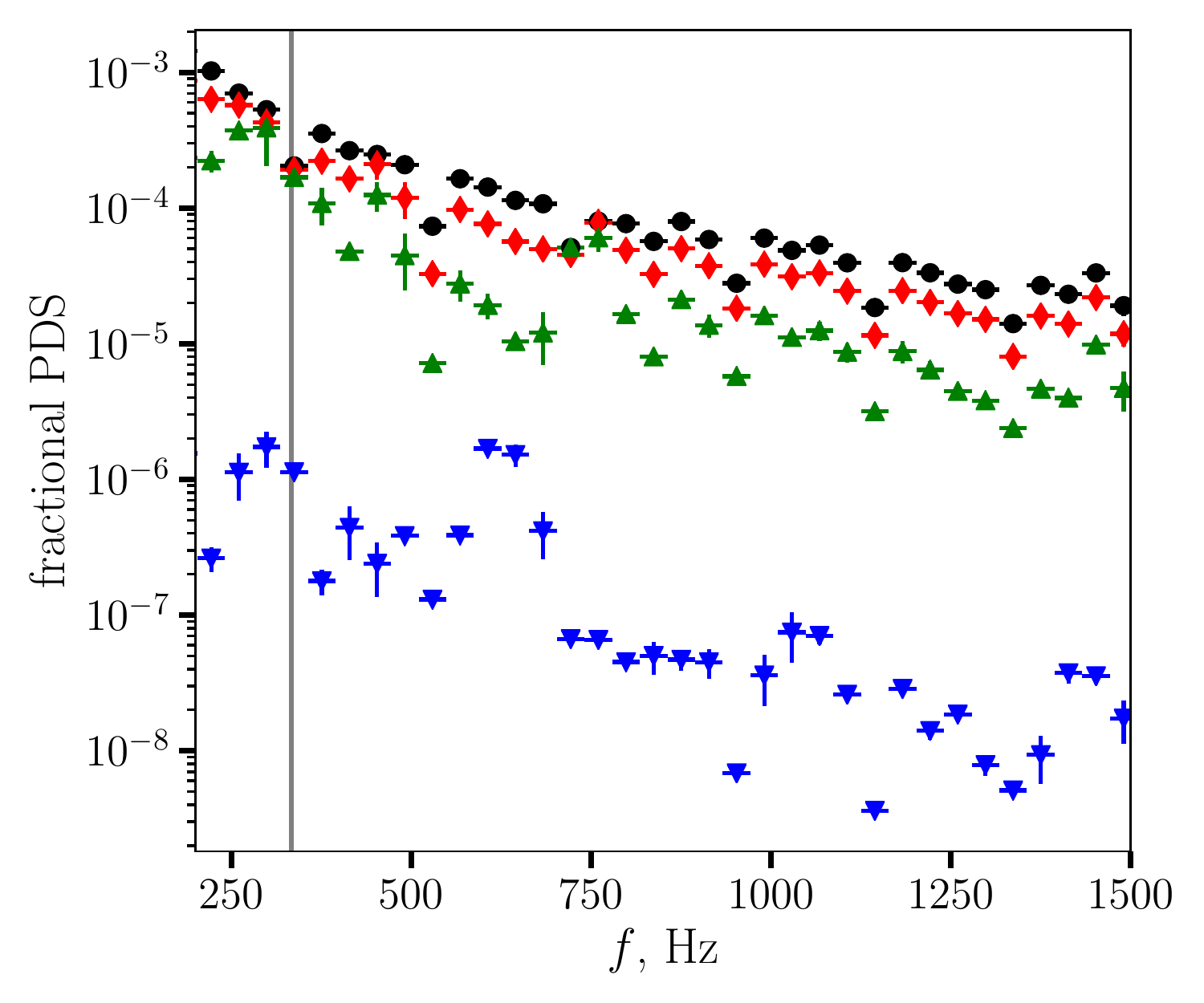}
\caption{Integral PDS (time span from $0.27$ to $0.32\,$s) for the simulation {\tt 8LR} for different inclinations (black circles $i_{\rm obs} =0$, red diamonds $i_{\rm obs}=45\degr$, green upward triangles $i_{\rm obs} = 90\degr$, and blue downward triangles $i_{\rm obs}=180\degr$). 
}\label{fig:pds4}       
\end{figure}

For the realistic mass-accretion-rate configuration, oscillations appear simultaneously with the development of the heating instability, and their contribution is  only clearly visible at particular inclinations (see Fig.~\ref{fig:pds4}). The two poles behave in a profoundly different way because of the asymmetry formed by heating instability. 
Further simulations of the {\tt 8LR} case are difficult because of the very strong density contrasts formed during the instability. 

\section{Discussion}\label{sec:disc}

In the dynamical spectra presented in Sect.~\ref{sec:res:lcurves}, there are clearly at least two types of quasi-periodic variability signals: one disappears for a pole-on observer and is thus related to non-axisymmetric structures (waves and vortices produced by shear instabilities); and the other is present at all inclinations. The frequency of this mode clearly increases with the flux, approximately as $f_{\rm QPO} \propto L_{\rm obs}^{1/3}$ (see Fig.~\ref{fig:ffreq3}). We leave a detailed study of the properties of the predicted QPOs to a separate paper. 

\begin{figure}
\includegraphics[width=\columnwidth]{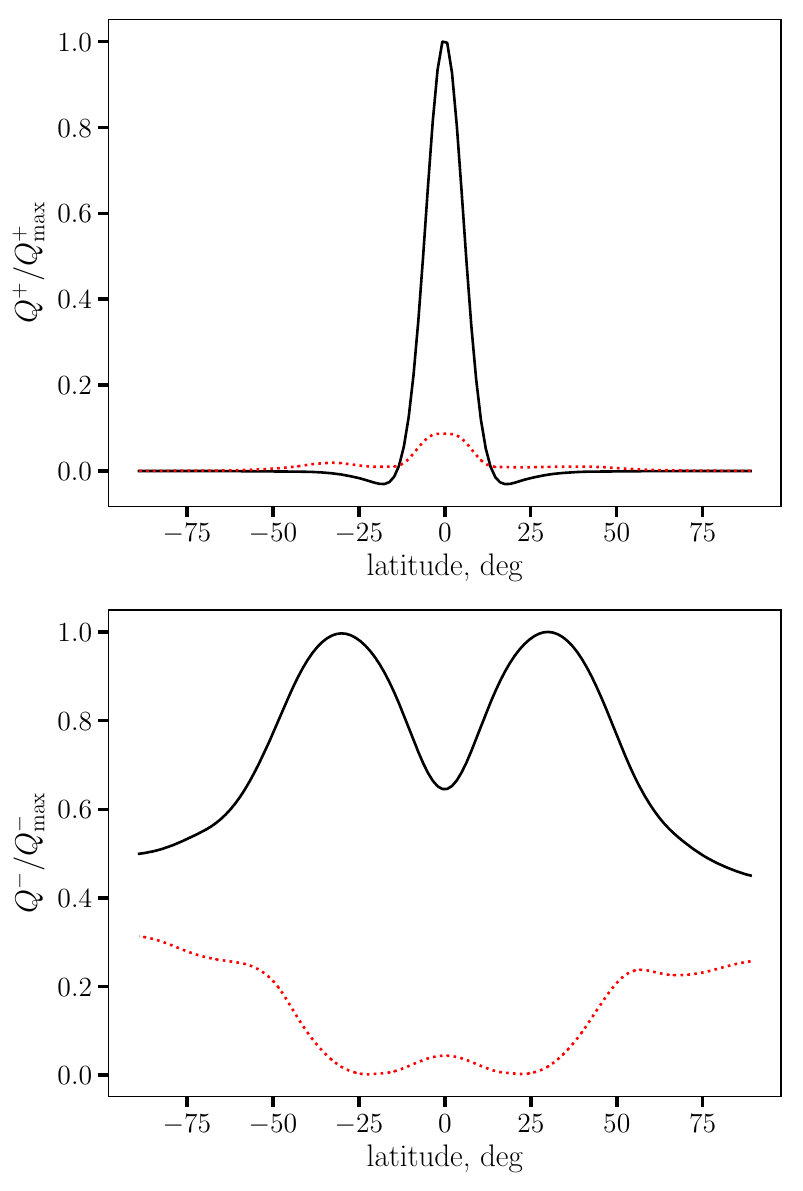}
\caption{Time- and longitude-averaged energy dissipation (upper panel) and radiation flux (lower panel) for the high-accretion-rate {\tt 3LR} simulation run. Mean values and root-mean-square deviations are shown, respectively, with black solid and red dotted curves. In each panel, the relevant quantity is normalised by the maximal averaged value.
}\label{fig:qpm}       
\end{figure}

It is natural to interpret this oscillation mode as a surface mode existing within the SL, as was done by \citet{PB04b} for dwarf-nova oscillations (DNOs). However, the width of the SL, independently from the assumptions, should increase with mass accretion rate. This is especially true for the radiation-pressure-supported case where the radiative flux is fixed by equilibrium with effective gravity $F= cg_{\rm eff}/\varkappa$, and the growth of the area over which the dissipation is spread should reflect the growth of the mass accretion rate. Approximately, the width of the SL grows linearly with the mass accretion rate.
As the speed of sound depends weakly on the mass accretion rate, the frequency of a DNO-like sonic mode for a thin radiation-pressure-supported SL is
\begin{equation}
    f_{\rm s} \sim \frac{1}{2\uppi} \frac{c_{\rm s}}{H} \propto \dot{M}^{-1}, 
\end{equation}
where $c_{\rm s}\sim \sqrt{p_0/\rho_0}$ is the speed of sound. 
However, this approach  assumes that the observed oscillations are produced near the equator. The equatorial belt is indeed responsible for most of the energy dissipation, but the radiation flux is  broadly distributed over the surface due to the importance of radiation
pressure, and most of the variability comes from high latitudes (see Fig.~\ref{fig:qpm}). This is visible in Fig.~\ref{fig:qpm}, where most of the dissipation in the system takes place directly at the equator; most of the radiation flux comes from intermediate latitudes ($30\div 40\degr$), while the most variable regions are at higher latitudes.
In addition, the SL itself cannot work as a proper waveguide because of the very strong velocity shear that exceeds the Keplerian frequency.

\begin{figure}
\includegraphics[width=\columnwidth]{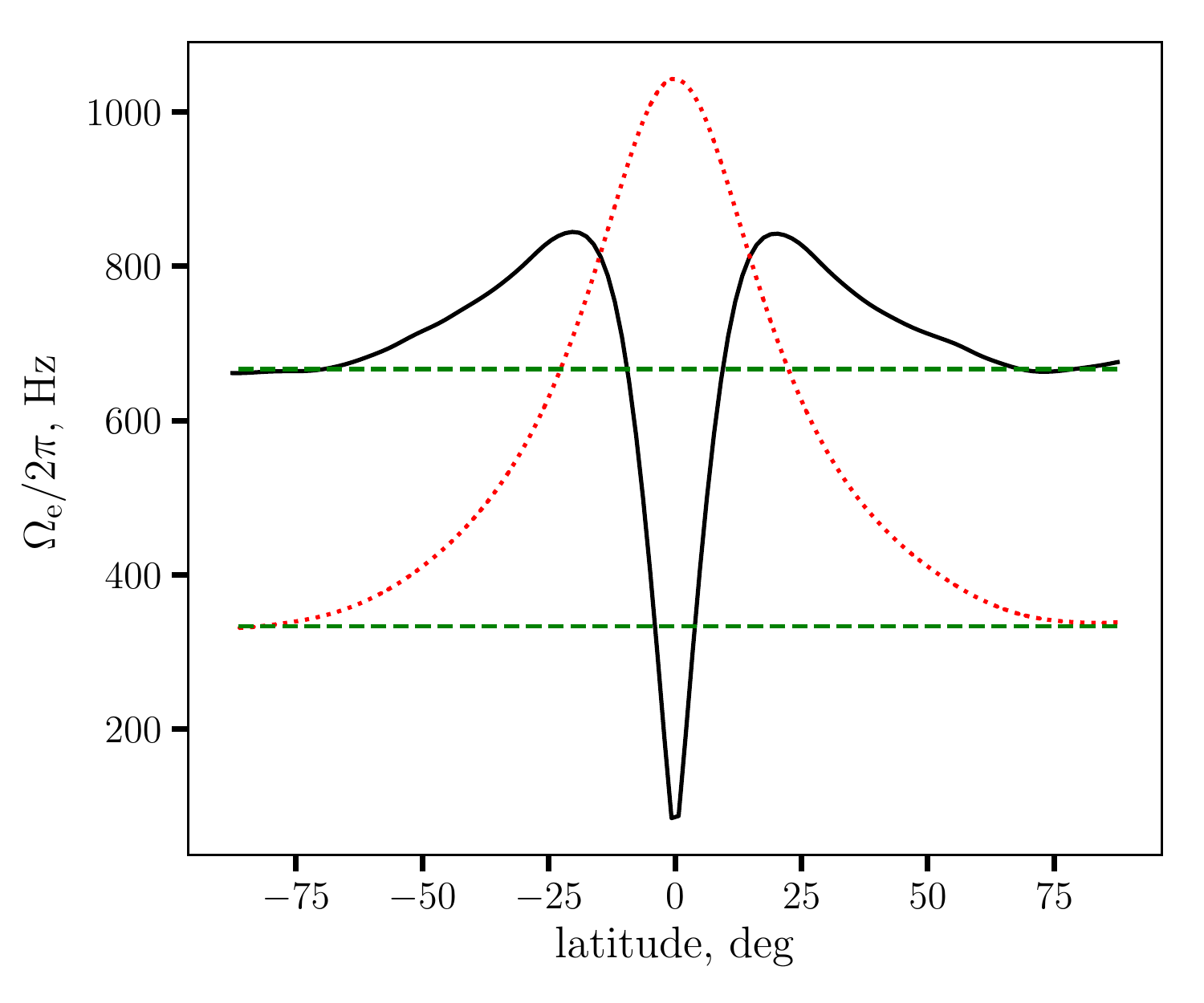}
\caption{Epicyclic (black solid) and local rotation (red dotted) frequencies for the model {\tt 3LR}, averaged in time between $0.3$ and $0.5\,$s and over the azimuthal angle. Green dashed horizontal lines correspond to the rotation rate of the NS and its second harmonic. 
}\label{fig:ekappa}       
\end{figure}

We can only conclude that the oscillations present in the observational data and in simulations are not the resonance frequencies for sonic waves but rather correspond to a different type of oscillation. The best candidate for these oscillation modes are r-modes (i.e. Rossby waves). 
As we show in Appendix~\ref{sec:app:ekappa},
their frequencies at a given co-latitude $\theta$ form an equidistant spectrum with
\begin{equation}
    f_{r, m} = \frac{1}{2\uppi}\left( \Omega_{\rm e} + m \Omega\right),
\end{equation}
where 
\begin{equation}
    \Omega_{\rm e} = \sqrt{2\Omega \ppardir{\theta}{\Omega\sin^2\theta }}
\end{equation}
is the epicyclic frequency (the frequency at which a portion of matter conserving its angular momentum and  only affected by gravity and inertial forces would oscillate in a latitudinal direction), $\Omega = \Omega(\theta)$ is the rotation frequency, and $m$ is a whole number. For rigid-body rotation, $\Omega_{\rm e} \simeq 2\Omega \cos\theta$. 
If the variability is excited in a slowly rotating region outside the SL itself, and the epicyclic frequency changes slowly throughout this region, we see one peak corresponding to the non-rotating $m=0$ mode at $f\lesssim 2f_{\rm spin}$ and aliases at frequencies differing by $\Delta f \simeq f_{\rm spin}$. This is similar to the spectra obtained in our simulations (see Sect.~\ref{sec:res:lcurves}), and at the same time similar to the pair of QPOs in LMXBs. 
As seen in Fig.~\ref{fig:ekappa}, there is a maximum of epicyclic frequency roughly in the interaction region between the SL and the slowly rotating matter. In addition, the epicyclic frequency in this region is very close to the local rotation frequency, meaning that the perturbations are in resonance with rotation. 
As most variability comes from higher latitudes, we propose that the oscillations are excited at intermediate latitudes ($30-50\degr$ for {\tt 3LR}), probably by shear instabilities in the interaction region, and then propagate towards the poles.

Let us assume that the oscillations are always generated at the latitude of the rim of the SL, all the energy is dissipated within the layer, and the local flux is equal to the Eddington flux $cg_{\rm eff}/\varkappa$. Flux scaling with the Eddington limit means that the luminosity should grow approximately linearly with the surface area of the layer, $L \simeq L_{\rm Edd} \cos\theta_{\rm SL}$. 
The epicyclic frequency should therefore scale as
\begin{equation}\label{E:disc:rledd}
    f_{\rm r0} \simeq 2f_{\rm spin} \theta_{\rm SL} \simeq 2 f_{\rm spin} \frac{L}{L_{\rm Edd}},
\end{equation}
which  reproduces the characteristic values of the frequency in our simulations but somewhat over-estimates the dependence on flux. As the radiating region does not exactly coincide with the SL, and the width of the SL also depends on the parameters of the inflow (the extreme case being the case of a strongly inclined source; see Eq.~\ref{E:lobs:inc}), and $g_{\rm eff}$ is generally smaller than gravity, we expect the linear scaling to be a very crude approximation, over-predicting the slope of the actual (seen in simulations) frequency dependence on flux. 

A similar type of QPO spectrum consisting of the local epicyclic frequency and its aliases with the rotation frequency was obtained by \citet{erkut08} and \citet{belyaev17} who considered a BL as a part of the accretion disc. However, in the case considered in these papers, both $\Omega_{\rm e}$ and $\Omega$ are close to the local Keplerian frequency and are not supposed to be sensitive to the spin of the NS.
\citet{belyaev17} also studied the conditions for the excitation of the oscillation modes, which are to some degree also applicable to our results, as the existence of a strong velocity shear in combination with differential rotation is a universal feature of any BL model. 
Shear instabilities may be excited without an initial velocity discontinuity, but the spatial scales of velocity variations should be smaller than the size of the simulation domain  \citep{BR12}, and the velocity profile should have an inflection point \citep{HK15}.
However, neither this model nor our simulations are so far capable of explaining clearly why only two peaks are observed in the PDSs of real LMXBs (though see above, Sect.~\ref{sec:res:lcurves}), and why, in a large number of LMXBs, the distance between the peaks is actually half of the spin frequency. 
Explaining and predicting the details of QPO features in the PDS requires more profound studies, both analytical and numerical. 
Both classical and spreading layer approaches have their limitations, as the real motions are likely to be three-dimensional \citep{bbp08}. 

The amplitudes of the oscillations observed in our simulations are rather small, of the order $10^{-4}$, but nevertheless the peaks themselves are clearly significant. This is much smaller than the observed $\sim 10\%$ \citep{mendez01}, probably due to several reasons. First, the observational data provide us with spectral variability that does not always follow the variations of the bolometric flux. It appears that the large amplitudes of the kHz QPOs in harder X-rays ($E\gtrsim 5\,\mbox{keV}$) are related to the temperature variations of the radiating surface, converted in the blackbody approximation to exponentially strong monochromatic flux variations. Still, the temperature variations required to reproduce the observed flux variability is several per cent. 
Another reason could be related to the algorithm we use to calculate the observables: it does not take into account either relativistic effects or the shape of the photosphere of the SL. In some of our models, the vertical thickness of the layer reaches hundreds of meters. The kilometer-scale variations of the shape of the photosphere potentially have important implications for the observed variability of BLs. Last but probably most important, we cannot exclude that the oscillations generated in the SL resonate or are amplified elsewhere, such as for example in an optically thin hot corona or in the accretion disc. This suggestion, though speculative, can help us to understand why only two harmonics are normally seen: structures more extended than BLs are unlikely to have resonance frequencies higher than $\sim 1.5\rm\,kHz$.

\section{Conclusions}\label{sec:conc}

In this paper, we consider a time-dependent hydrodynamic SL on the surface of a NS. We used two-dimensional spectral modelling to resolve the evolution of the differentially rotating flow. 
We find that, though challenging due to the super-sonic compressible nature of the flow, spectral simulations of a SL on the surface of a NS may be quite productive. 
We mainly consider the interaction of a new material rotating close to Keplerian velocity with the old, spun-down atmosphere of the NS, and this interaction produces a set of hydrodynamical phenomena that have a huge impact on the dynamics of the system. In particular, the velocity shear is susceptible to shear instability modes that provide angular momentum transfer within the layer and excite inertial oscillation modes closer to the poles where they produce variability patterns closely resembling kHz QPOs in real LMXB systems. 

\begin{acknowledgements}
We would like to thank Alexander Philippov for his help with the spectral filtering, and Roman Rafikov for discussions about different oscillation modes. 
We also thank Yuri Levin, Andrei Beloborodov, Andrei Gruzinov, Lorenzo Sironi, Joe Patterson, and Anatoly Spitkovsky for discussions on the different aspects of boundary layer physics and the referee for numerous valuable comments. 
PA acknowledges the support from the Program of development of M.V. Lomonosov Moscow State University (Leading Scientific School `Physics of stars, relativistic objects and galaxies'). 
JP was supported by the grant 14.W03.31.0021 of the Ministry of Science and Higher Education of the Russian Federation.
We acknowledge the use of the Finnish Grid and Cloud Infrastructure and the Finnish IT Center for Science (CSC).  
\end{acknowledgements}

\bibliographystyle{aa}
\bibliography{mybib}

\onecolumn

\begin{appendix}

\section{Derivation of the equations for divergence and vorticity}\label{sec:derive}

To get the equations for $\delta = \nabla \cdot \vector{v}$ and $\omega = \left(\nabla \times \vector{v}\right)_r$, we need to take divergence and curl of the system of dynamical equations containing sources and sinks
\begin{equation}\label{E:app:euler}
    \pardir{t}{\vector{v}} + \left(\vector{v}\cdot \nabla\right)\vector{v} = - \frac{1}{\rho} \nabla p + \vector{g} +\frac{\rho^+}{\rho} \vector{v}_{\rm source}- \frac{\vector{f}_{\rm fric}}{\rho},
\end{equation}
where $\vector{g} =  -\nabla \Phi$ is gravity without centrifugal terms, and then integrate them in radial direction. Gravitational potential has the form $\displaystyle\Phi = -\frac{GM}{R} + \Delta\Phi$, where $\Delta \Phi$ accounts for a non-spherical shape of the star and depends on the latitude and longitude. 
For the friction force $\vector{f}_{\rm fric}$, we use the form $\vector{f}_{\rm fric} = - \frac{1}{t_{\rm fric}} \left( \vector{v}-\vector{v}_{\rm NS}\right)$. 
Here, $v_{\rm NS} = \Omega_* R_* \sin\theta$ is the rotation velocity field of the NS itself  and is directed azimuthally. 

Before deriving the actual equations for vorticity and divergence, we note that, for any velocity field, its vector product with its curl $\vector{\omega} = \nabla \times \vector{v}$ is 
\begin{equation}\label{E:app:vcrossomega}
    \vector{v} \times \vector{\omega} = \frac{1}{2}\nabla^2\vector{v} - (\vector{v}\nabla) \vector{v}.
\end{equation}
Hence, 
\begin{equation}\label{E:app:dvcrossomega}
    \nabla \times \left(\vector{v} \times \vector{\omega}\right) = - \nabla \times  (\vector{v}\nabla) \vector{v},
\end{equation}
and the curl of the non-linear term $(\vector{v}\nabla)\vector{v}$ in Eq.~(\ref{E:app:euler}), after application of the triple vector product rule, takes the form
\begin{equation}
    \nabla \times  (\vector{v}\nabla) \vector{v} = - \nabla \times \left(\vector{v} \times \vector{\omega}\right) = \delta \, \vector{\omega} + (\vector{v}\nabla ) \vector{\omega} - \vector{v} (\nabla \cdot \vector{\omega}) - (\vector{\omega} \nabla) \vector{v}.
\end{equation}
As $\vector{\omega}$ has zero divergence, the curl of Eq.~(\ref{E:app:euler}) becomes
\begin{equation}\label{E:app:curl}
    \displaystyle   \pardir{t}{\vector{\omega}} +
  (\vector{v}\cdot \nabla)\vector{\omega} = (\vector{\omega} \cdot
\nabla) \vector{v} - \delta \, \vector{\omega} +
  \frac{1}{\rho^2}\nabla p \times \nabla \rho -  \nabla \times \left( \frac{\rho^+}{\rho} \vector{v}_{\rm d} \right)- \nabla \times \frac{\vector{f}_{\rm fric}}{\rho},
\end{equation}
where $\omega_{\rm d} = 2c_{\rm K}\Omega_{\rm K} \cos\alpha $ is the vorticity in the source of matter (see Sect.~\ref{sec:vert}).
Taking radial integral of the radial component of Eq.~(\ref{E:app:curl}) is straightforward, as the equation does not contain radial derivatives. Finally, we get Eq.~(\ref{E:euler:omega}):
\begin{equation}\label{E:app:omega}
  \displaystyle  \pardir{t}{\omega} + \nabla \cdot (\omega \vector{v}) =
  -\nabla\times \frac{\nabla \Pi}{\Sigma}  
  \displaystyle  +\left( \omega_{\rm d}-\omega\right) \frac{S^+}{\Sigma} +
\left[ (\vector{v}_{\rm d}-\vector{v})\times \nabla \frac{S^+}{\Sigma}\right]_r
  + \frac{1}{t_{\rm fric}} \left( \Omega_*
  -\omega\right).  
\end{equation}
 The right-hand side of this equation contains a baroclinic term capable of creating vorticity out of density and pressure variations, and three terms related to the vorticity of the accreted matter $\omega_{\rm source}$ and friction with the surface. 

Another equation describing the time evolution of $\delta$ comes from taking divergence of dynamical equation. It is rather non-trivial to expand the advection term, $\nabla \cdot ((\vector{v}\cdot \nabla)\vector{v})$. Therefore, let us first note that, according to Eq.~(\ref{E:app:vcrossomega}),
\begin{equation}\label{E:app:vomega}
\displaystyle \nabla \cdot \left( \vector{v} \times \vector{\omega}\right) = 
 \nabla^2\frac{v^2}{2} - \nabla \cdot \left((\vector{v}\cdot \nabla)\vector{v}\right).
\end{equation}
Here, the last term on the right-hand side is identical to the advective left-hand-side term in the derivative of Eq.~(\ref{E:app:euler}), hence
\begin{equation}\label{E:app:delta}
\pardir{t}{\delta} = \nabla \cdot \left( \vector{v}\times 
  \vector{\omega}\right) -  \nabla \cdot \left(\frac{\nabla \Pi}{\Sigma}\right)
  - \nabla^2 \left( \frac{v^2}{2} + \Delta \Phi \right)
  - \delta \frac{S^+}{\Sigma} + \left( \vector{v}_{\rm d} - \vector{v}\right) \nabla \frac{S^+}{\Sigma} - \frac{\delta}{t_{\rm fric}}.
\end{equation}
The potential term $\Delta \Phi$ appears because of rotational deformation of the star. As the surface of the star should be an equipotential surface, total gravitational and centrifugal potential 
\begin{equation}
    \Phi - \frac{1}{2}R^2\Omega_*^2  = -\frac{GM}{R} + \Delta \Phi - \frac{1}{2}R^2\Omega_*^2 = \mbox{const},
\end{equation}
implying $\Delta \Phi = \frac{1}{2}R^2 \Omega_*^2$. 

\section{Numerical implementation and tests}\label{sec:numerics}

\subsection{Code description}

The problem we consider is essentially a more physically elaborate version of
shallow-water hydrodynamics, supplemented with energy transfer and source and
sink terms. As the problem is formulated for the surface of a sphere, it is
natural to use a spectral code working with spherical harmonics. We
used the {\tt shtns} library \citep{shtns} designed for hydrodynamical and
geophysical applications\footnote{We also used a wrapper class {\tt spharmt} (\url{https://gist.github.com/jswhit/3845307}) for {\tt shtns} quantities and operators written by Jeffrey Whitaker.}.
A major challenge of our problem is that, unlike the classical shallow-water
physics, it is far from the subsonic Rossby-approximation motions, and thus the
time step is limited by several processes. 
One of the requirements for the time step is the Courant-Friedrichs-Lewy condition \citep{CFL} of the form
\begin{equation}\label{E:dtCFL}
\Delta t \leq \Delta t_{\rm CFL} = \frac{C}{u} \Delta x_{\rm min},
\end{equation}
where $\Delta x_{\rm min}$ is the minimal physical size of a cell in the
simulation domain, $u$ is the fastest relevant signal propagation velocity,
and $C\lesssim 1$ is a constant related to the particular
solver used in simulations. The existence of sources and sinks for several
physical quantities sets additional upper limits for the time step and requires
us to adjust the time step with the physical conditions. We compute the actual
time step as a harmonic sum of several time steps
\begin{equation}\label{E:dt}
\displaystyle \Delta t = \left( \sqrt{C_{\rm sound}c^2_{\rm s, \,max}+C_{\rm adv}v^2_{\rm max}} \Delta x_{\rm min}^{-1}
+ C_{\rm thermal}\Delta t_{\rm thermal}^{-1} + C_{\rm accr}\Delta t_{\rm accr}^{-1} \right)^{-1},
\end{equation}
where $C_{\rm sound, \, adv, \,thermal, \,accr} \gtrsim 1$ are dimensionless adjustable parameters regulating the role of each time step (note that a larger coefficient here means a smaller time step). Thermal, $\Delta t_{\rm thermal} = \min{E/(Q^+ +Q^-),}$
and accretion, $\Delta t_{\rm accr} = \min{\Sigma/(S^++S^-),}$ time steps are estimated as the time steps required to resolve temporally thermal and accretion processes, respectively. 
Such a choice for a variable time step ensures that all the relevant physical processes can be resolved: sonic wave propagation, dissipation, radiation losses, and accretion. 

As spectral methods tend to produce high-frequency noise, diffusion-like dissipation terms were introduced for all the five principal quantities $\omega$, $\delta$, $\Sigma$, $E$, and $a$.
This is done for the numerical implementation of Eqs.~(\ref{E:mass:sigma}),
(\ref{E:euler:omega}), (\ref{E:euler:delta}), and (\ref{E:energy:Eint}).
In spectral space, an additional dissipation term may be viewed as a multiplier cutting off high frequencies (low-pass filter). A usual form for this low-pass filter (see e.g. \citealt{phaedra}) is hyperdiffusion
\begin{equation}\label{E:dissp}
\displaystyle d_{\rm HD}(l) = 
e^{-\displaystyle\frac{\Delta t}{t_{\rm D}}  \left(\frac{L}{L_{\rm max}}\right)^{N_{\rm diss}/2} 
},
\end{equation}
where $L=- l(l+1)/R_*^2$ is the Laplacian acting on the spherical harmonic of degree $l$ (see \citealt{mathphys}),  $L_{\rm max}$ is its maximal value (corresponding to the grid resolution), $t_{\rm D}$ is the characteristic dissipation timescale on the size of the grid, and  $N_{\rm diss} \geq 2$ is a free parameter describing the shape of the low-pass filter. 
If $N_{\rm diss} = 2$, the filtering procedure is equivalent to a regular diffusion term added to the right-hand sides of all the main differential equations.
The flow as a whole is negligibly affected if the dissipation factor for the lowest-order harmonic is indistinguishable from zero $t_{\rm D} = \gtrsim |\ln e_{\rm M}| \Delta t$ \citep{phaedra}, where $e_{\rm M}$ is machine precision ($e_{\rm M} \simeq 2\times 10^{-16}$ in all the simulations we run). 

Effectively, filtering with $N_{\rm diss} >2$, while more efficient than
normal diffusion, cuts all the high frequencies  very sharply. 
Such spectral truncation leads to its own noise, especially close to 
discontinuities. This is known as Gibbs phenomenon \citep{gibbs} and leads to
oscillations that may be amplified by the non-linear nature of our system of
equations.
For $\Sigma$ and $E$ that can vary sharply and span several
orders of magnitude, but become unphysical if negative, the Gibbs phenomenon may
become disastrous. 
A reasonable solution introducing the little Gibbs effect and providing an extremely accurate treatment to the large-scale flow is
\begin{equation}\label{E:dissp:decosine}
\displaystyle d(l) = e^{ - \displaystyle \frac{L}{L_{\rm max}} \left(\frac{L}{L_{\rm max}}+\frac{1}{R_*^2L_{\rm max}}\right)  \frac{\Delta t}{t_{\rm D}} },
\end{equation}
which we use for all the simulations in this paper.

The energy lost by the flow is added as a source
of internal energy, as described in Sect.~\ref{sec:energy}.
By adding dissipation as a source of energy, we are likely to introduce
high-frequency noise to the energy field, and therefore the dissipation field was
smoothed in the same way as the basic quantities (see Eq.~\ref{E:dissp}), but with a shorter diffusion timescale.
The exact value of the dissipation smoothing parameter affects the thermal stability of the simulation but does not change the overall dynamics. From the physical point of view, it only ensures that the dissipation does not significantly vary within a single resolution element. 

The code itself is written as a hybrid \textsc{Python3}/\textsc{C++} program. 
All the numerically heavy (spectral) calculations are solved with the C++ \textsc{shtns} library, whereas the main loop and the related high-level functionalities are operated from the more user-friendly \textsc{Python3} driver.
In our experience, this provides a good balance between numerical efficiency and ease of use.
The spherical harmonic calculations are parallelised using the shared memory paradigm with \textsc{openMP} pragmas. 
This enables us to take advantage of multi-core platforms ranging from powerful desktop computers to occupying one complete node in computing clusters.
The Hierarchical Data Format (HDF5) is used to save and store the simulation results. 

\subsection{Tests}

In Table~\ref{tab:tests}, we list the test models we calculated with their basic parameters. 

\subsubsection{Zero-accretion-rate, rigid-body rotation case}\label{sec:tests:ND}
 
 As one of the tests, we try evolution of a layer with initial surface density $\Sigma_0 =10^8\gcm$ and sound velocity $c_{\rm s} \simeq 1.7\times 10^{-3}c$ without accretion or depletion (test models {\tt NDLR/NDHR}). As in all the other models, an initial perturbance of 5\% was introduced. The NS spin period was set to $3\,$ms. The Mach number of this flow is about 50. 
 For this test, we also turn off dissipation heating and radiation losses.
 Without thermal effects, rotation profile in such a model should not change with time, and the perturbation proceeds rotating with the surface of the star. 
 To check the accuracy of this solution, it is sufficient to correct for the rotation angle, interpolate from one grid to another, and estimate the standard deviation or the maximal deviation between the map calculated by the code and the interpolated initial map.
 
\begin{figure}
\includegraphics[width=\columnwidth]{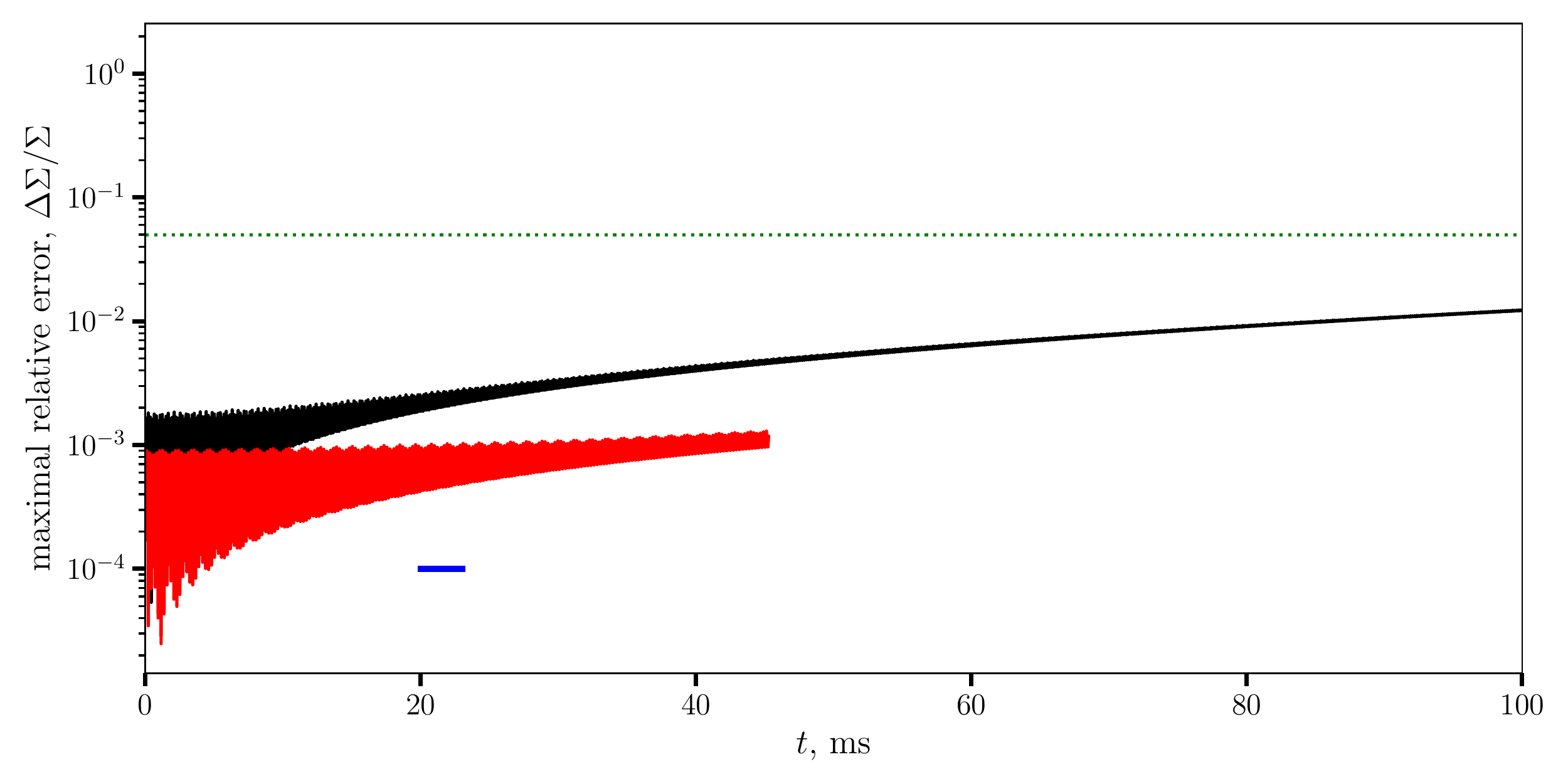}
 \caption{Maximal relative error in surface density $\max \frac{\Delta \Sigma}{\Sigma}$ for the test models {\tt NDLR} (black) and {\tt NDHR} (red). The blue horizontal segment in the lower panel has the length of one spin period. The dotted green horizontal line corresponds to the amplitude of the initial perturbation, $0.05$.}\label{fig:rtests}       
\end{figure}

Figure~\ref{fig:rtests} shows how the maximal relative difference in surface density evolves with time. 
The errors around $10^{-3}$ are interpolation errors.
As we can see, supersonic rotation is reasonably well tracked for multiple rotation periods, and the accuracy is better for a finer grid.

\subsubsection{Split-sphere tests}\label{sec:tests:twist}

\begin{figure*}
\includegraphics[width=\textwidth]{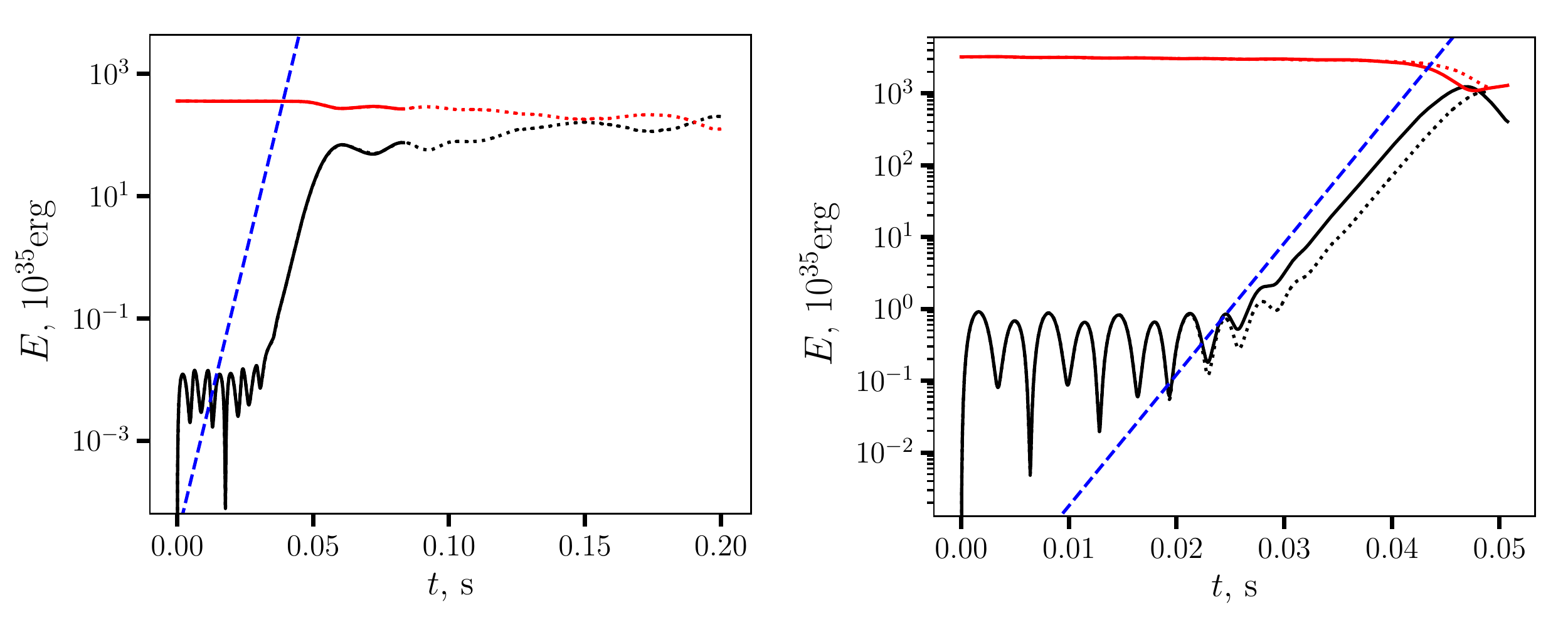}
 \caption{Energy evolution and relaxation for the split-sphere test, sub- (left panel) and super-sonic (right panel) cases. Black lines correspond to the part of kinetic energy related to $v_\theta$, red to $v_\varphi$.
 Dotted lines are used for lower-resolution models {\tt (s)twistLR}, solid lines for high-resolution {\tt (s)twistHR}. 
 Blue dashed lines show an exponential law $\propto e^{\Omega_* t}$.
 }\label{fig:KH}       
\end{figure*}

\begin{figure*}
\includegraphics[width=\textwidth]{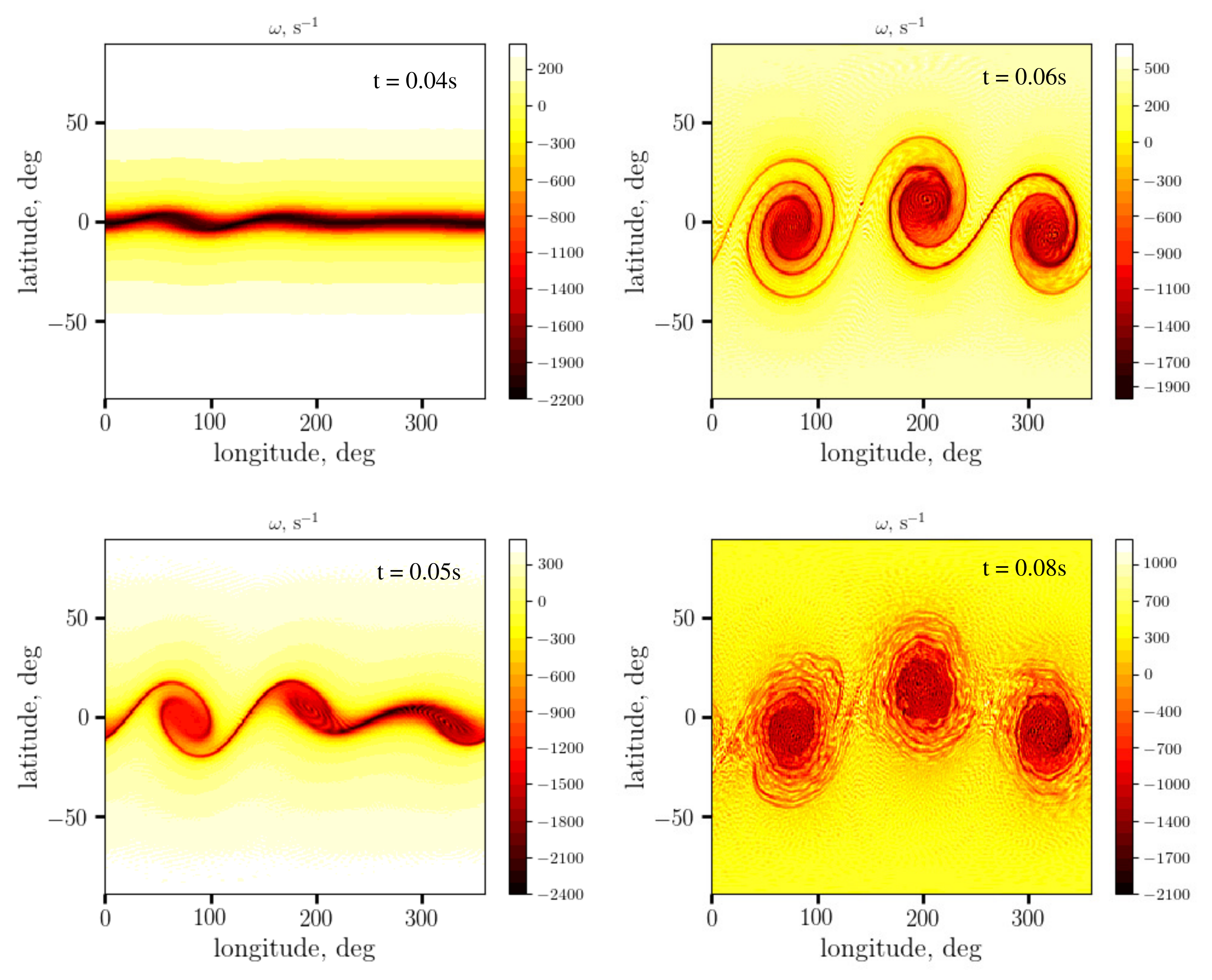}
\caption{Four vorticity snapshots ($t = 0.04,0.05, 0.06,$ and $0.08$s) of the Kelvin-Helmholtz instability development in the split-sphere simulation, model {\tt twistHR}.
 }\label{fig:KH:snapshots}       
\end{figure*}

The purpose of this test set ({\tt twistLR}, {\tt twistHR}, {\tt stwistLR}, and
{\tt stwistHR}) was to trace the development of sub- and super-sonic shear
instabilities on a sphere. Rigid-body rotation ($P_{\rm
  spin}=10$ and $30\,$ms) was modified by a factor rapidly changing from $-1$ to $1$ near the equator
\begin{equation}
\displaystyle  \Omega = \Omega_* \frac{\uppi/2 - \theta}{\sqrt{(\uppi/2-\theta)^2+\Delta\theta^2}},
\end{equation}
where $\Delta\theta$ was set to 0.1 for all the models. The choice of the effective temperature (set by $Q_{\rm NS}$, see Sect.~\ref{sec:energy}) makes some simulations subsonic and others supersonic. For
$\Delta\theta \ll \uppi$, subsonic configuration is unstable to Kelvin-Helmholtz
instability. Instability at high wavenumbers is suppressed by the finite
shear value \citep{ray82}, hence the fastest-growing unstable modes are two- and three-armed, with the increment of about $\Omega_*$ (see Fig.~\ref{fig:KH}). 
The sharper the velocity gradient, the higher the fastest-growing mode. 
Conservation of angular momentum prevents the formation of a single vortex.  
The primary instability mode changes the overall
velocity field into a set of vortices centered in the initial equatorial region. Vorticity evolution during the instability development phase is
shown in Fig.~\ref{fig:KH:snapshots}. 

\begin{figure*}
\includegraphics[width=\columnwidth]{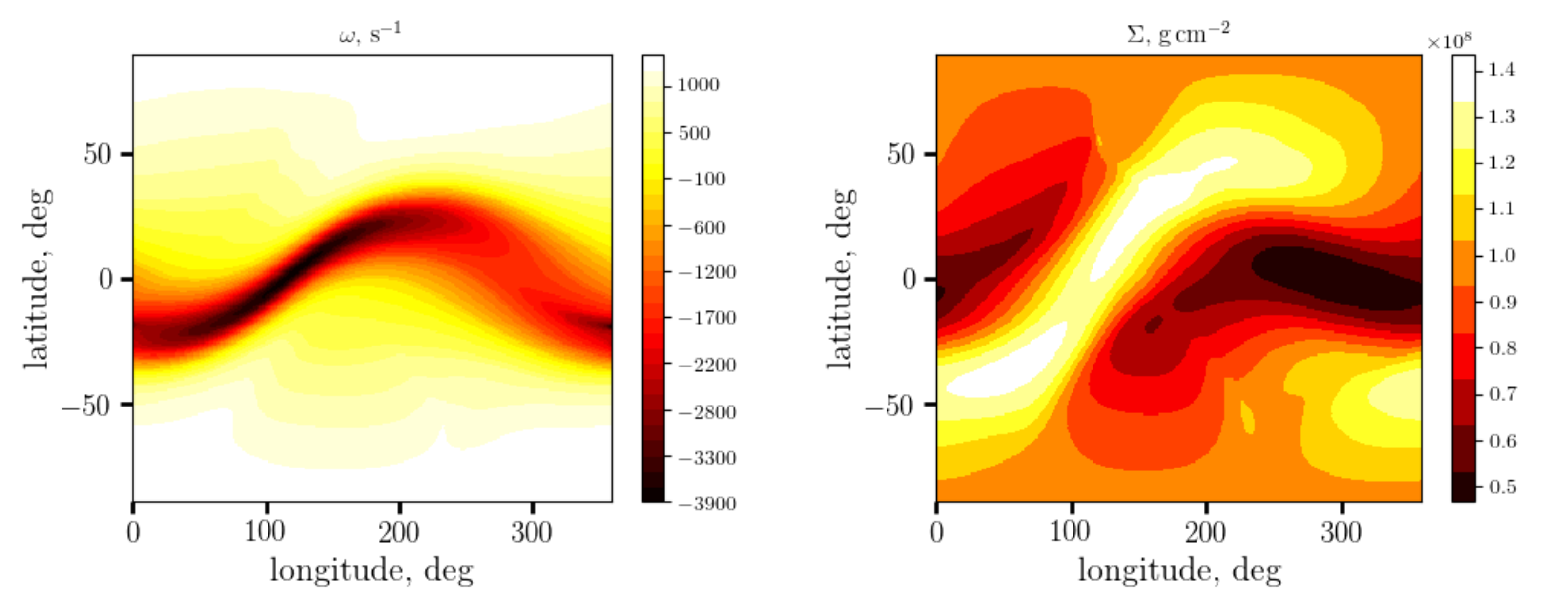}
\caption{Snapshot of the {\tt stwistLR} simulation at $t=0.04\,$s, when the instability is fully developed and evolves into a system of shock waves. Vorticity is shown in the left and the surface density in the right panel.}
\label{fig:KH:stwist}       
\end{figure*}

For the mildly supersonic split-sphere test ($P_{\rm spin} = 10\,$ms), a supersonic shear instability develops on similar timescales close to the rotation period is used as the basis for the split-sphere rotation. 
However, instead of vortices, a system of standing shock waves is formed (see Fig.~\ref{fig:KH:stwist}).

As we can see, development of shear instabilities conforms well with the expectations based on analytical and numerical studies of the subject. First, there is dynamical-timescale exponential growth of the instability, subsequently evolving into an equipartition turbulent stage. For the subsonic case, numerical resolution does not affect the results considerably during the time span of the calculations. 

\begin{table*}\centering
\caption{Test simulations.}
\label{tab:tests}
\bigskip
\begin{tabular}{lccccccc}
\hline
Model ID & dimensions & $t_{\rm max}$ & $\Sigma_{\rm init}$ & $P_{\rm spin}$  & comments\\
  &   &   s &   \gcm &   ms &  \\
\hline
NDLR & 128$\times$256 & 0.1 & $10^8$ &   3&  \\
NDHR & 256$\times$516 & 0.04 & $10^8$ & 3&  \\
twistLR  & 128$\times$256 & 0.15 & $10^8$ & 30& subsonic twist test \\
twistHR  & 256$\times$512 & 0.15 & $10^8$ & 30& subsonic twist test \\
stwistLR & 128$\times$256 & 0.04     &  $10^8$ &  10  &  supersonic twist test \\
stwistHR & 256$\times$512 & 0.03     &  $10^8$ &   10 &  supersonic twist test \\
\hline
\end{tabular}
\end{table*}

\section{Frequencies of inertial modes on a differentially rotating sphere}\label{sec:app:ekappa}

Assuming an axisymmetric, differentially rotating velocity background, we linearise the set of dynamic equations and derive
a dispersion relation for small-amplitude shallow-water waves on a unit sphere. Perturbed quantities to be considered are density $\rho = \rho_0 +\delta \rho(\theta,\varphi, t)$, longitudinal velocity $v_\varphi = \Omega(\theta) \sin\theta + \delta v_\varphi(\theta,\varphi, t)$, and latitudinal velocity $v_\theta = \delta v_\theta(\theta,\varphi, t)$, where the terms with $\delta$ are small perturbations. The background flow is assumed to be pure differential rotation parametrized by angular velocity distribution $\Omega(\theta)$. All the perturbations are
expressed in exponential form $\propto \exp(\i(\omega t - k_\theta \theta - k_\varphi \varphi))$. 

First-order perturbation of the continuity equation in such assumptions is
\begin{equation}\label{E:WKB:continuity}
\displaystyle \left(\omega-k_\varphi \Omega\right) \frac{\delta \rho}{\rho} = k_\theta v_\theta +
\frac{k_\varphi \delta v_\varphi}{\sin \theta} .
\end{equation}
The two tangential Euler equations may, in general form, ignoring the terms containing radial velocities, be written as
\begin{equation}\label{E:WKB:Eulertheta}
\displaystyle \pardir{t}{v_\theta} + \frac{v_\varphi}{\sin \theta}
\pardir{\varphi}{v_\theta} - v_\varphi^2 \cot \theta = -\frac{1}{\rho}
\pardir{\theta}{p}
,\end{equation}
and
\begin{equation}\label{E:WKB:Eulerphi}
\displaystyle  \pardir{t}{v_\phi} + v_\theta \pardir{\theta}{v_\varphi}
  +\frac{v_\varphi}{\sin\theta}\pardir{\varphi}{v_\varphi}+ v_\varphi v_\theta
  \cot \theta = -\frac{1 }{\rho \sin \theta} \pardir{\varphi}{p}.
\end{equation}
For a super-sonic flow, contributions of the pressure variations on the right-hand side of the equations are of secondary importance, though in reality they are responsible for pressure and gravity oscillation modes. If we ignore the pressure variations, the two Euler equations become, respectively,
\begin{equation}\label{E:WKB:theta}
\displaystyle \tilde{\omega}^2  v_\theta= - 2 \i\, \Omega(\theta)\,
\tilde{\omega}\, \cos\theta \,\delta v_\varphi,
\end{equation}
and
\begin{equation}\label{E:WKB:phi}
\displaystyle
\tilde{\omega}^2\delta v_\varphi = \i\, \tilde{\omega}\, \ppardir{\theta}{\Omega(\theta) \sin^2\theta} \frac{v_\theta}{\sin \theta},
\end{equation}
where $\tilde{\omega} = \omega - k_\varphi \Omega$. 
Excluding the velocity components $v_\theta$ and $\delta v_\varphi$ from Eqs.~(\ref{E:WKB:phi}) and (\ref{E:WKB:theta}) yields a dispersion equation
\begin{equation}\label{E:WKB:deq}
\displaystyle \left(\tilde{\omega}^2  - \Omega_{\rm e}^2 \right)\tilde{\omega}^2 = 0,
\end{equation}
where
\begin{equation}\label{E:WKB:varkappa}
  \Omega_{\rm e}^2 = 2\Omega \cot\theta \ppardir{\theta}{\Omega \sin^2\theta}
\end{equation}
is the square of the local epicyclic frequency in the sense that a particle with a conserved angular momentum, confined to the surface of the star and being a subject of gravity and centrifugal force, will oscillate in latitudinal direction at this frequency. The possible values of $k_{\varphi}$ are restricted by the longitudinal periodic boundary conditions to be $k_\varphi = m$, where $m$ is a whole number.
Thus, the spectrum of possible inertial oscillation frequencies takes the form
\begin{equation}\label{E:WKB:inertial}
\displaystyle   \omega_{\rm inertial} = m\Omega \pm \Omega_{\rm e}.
\end{equation}
Without any loss of generality, we choose the sign in Eq.~(\ref{E:WKB:inertial}) to be plus. 
In the case of $m=0$ and $\Omega\simeq \Omega_*$, the only axisymmetric inertial mode has $\omega_{\rm inertial,\, 0} = \Omega_{\rm e} \simeq 2\Omega \cos\theta$ reproducing the Coriolis oscillation regime. Variability occurring in the regions co-rotating with  the NS would produce an equidistant spectrum of eigenmodes
\begin{equation}
    \omega_{\rm inertial,\, co-rotating} \simeq \Omega_{\rm e} + m \Omega_* ,
\end{equation}
with the frequencies differing by the rotation frequency of the star. 

\end{appendix}
\end{document}